\documentclass[aps,preprint]{revtex4-1}

\usepackage[T1]{fontenc}
\usepackage[latin1]{inputenc}
\usepackage{amsmath}
\usepackage[dvipdfmx]{graphicx}
\usepackage{amssymb}

\usepackage{color} 
\usepackage{graphicx}
\usepackage{bm}
\usepackage{amsmath}

\usepackage{here}

\makeatletter
\begin{document}

%
\def\average#1{\left\langle {#1} \right\rangle}
\def\averagefr#1{\left\langle {#1} \right\rangle_0}
\def\avbar#1{\wideoverline{#1}}
\def\impaverage#1{\left\langle {#1} \right\rangle_{\rm i}}
\def\bra#1{\left\langle{#1} \right|}
\def\ket#1{\left|{#1} \right\rangle}
\def\braket#1#2{\left.\left\langle{#1} \right|{#2}\right\rangle}
\def\ddelo#1{\frac{d^2}{d #1^2}}
\def\ddel#1#2{\frac{d^2 #1}{d #2 ^2}}
\def\ddelpo#1{\frac{\partial^2}{\partial #1^2}}
\def\delo#1{\frac{d}{d #1}}
\def\delpo#1{\frac{\partial}{\partial #1}}
\def\deldo#1{\frac{\delta}{\delta #1}}
\def\del#1#2{\frac{d #1}{d #2}}
\def\delp#1#2{\frac{\partial #1}{\partial #2}}
\def\deld#1#2{\frac{\delta #1}{\delta #2}}
\def\vvec#1{\stackrel{{\leftrightarrow}}{#1}} 
\def\vectw#1#2{\left(\begin{array}{c} #1 \\ #2 \end{array}\right)}
\def\vecth#1#2#3{\left(\begin{array}{c} #1 \\ #2 \\ #3 \end{array}\right)}
\def\mattw#1#2#3#4{\left(\begin{array}{cc} #1 & #2 \\ #3 & #4 \end{array}\right)}
\def\Eqref#1{Eq. (\ref{#1})}
\def\Eqsref#1#2{Eqs. (\ref{#1})(\ref{#2})}
\def\Eqrefj#1{(\ref{#1})式}
\def\Eqsrefj#1#2{\UTF{FFFD}\UTF{FFFD}?\ref{#1})(\ref{#2})}
\def\dispeq#1{$\displaystyle{#1}$}
\def\ret{{\rm r}}
\def\adv{{\rm a}}
\def\L{{\rm L}}
\def\R{{\rm R}}
\def\gf#1#2{g_{#1}^{#2}}
\def\jspin#1#2{j_{{\rm s}#1}^{#2}}
\def\jvspin#1#2{{\jv}_{{\rm s}#1}^{#2}}
\def\ls#1{\ell_{{\rm s}#1}}
\def\As#1#2{A_{{\rm s},#1}^{#2}}
\def\Es#1{E_{{\rm s},#1}}
\def\Bs#1{B_{{\rm s},#1}}

\renewcommand{\figurename}{Figure}

\def\listitem#1{\begin{itemize}\item #1 \end{itemize}}
\newcommand{\lt}{\left}
\newcommand{\rt}{\right}
\newcommand{\nablalr}{\stackrel{\leftrightarrow}{\nabla}}
\newcommand{\nablarl}{\stackrel{\leftrightarrow}{\nabla}}
\newcommand{\nablal}{\stackrel{\leftarrow}{\nabla}}
\newcommand{\nablar}{\stackrel{\rightarrow}{\nabla}}
\newcommand{\nnr}{\nonumber\\}
\newcommand{\adag}{{a^{\dagger}}}
\newcommand{\alphav}{{\bm \alpha}}
\newcommand{\alphaz}{\alpha_0}
\newcommand{\alphasf}{\alpha_\spinflip}
\newcommand{\alphaR}{{{\alpha}_{\rm R}}}
\newcommand{\alphaRv}{{\bm \alpha}_{\rm R}}
\newcommand{\Area}{A}
\newcommand{\aT}{\overline{\rm T}}
\newcommand{\av}{{\bm a}}
\newcommand{\Av}{{\bm A}}
\newcommand{\Asv}{{\bm A}_{\rm s}}
\newcommand{\Avem}{{\bm A}_{\rm em}}
\newcommand{\Aem}{{A}_{\rm em}}
\newcommand{\Ams}{{\rm A/m}^2}
\newcommand{\Aph}{A^{\phi}}
\newcommand{\Ath}{A^{\theta}}
\newcommand{\Aphv}{\Av^{\phi}}
\newcommand{\Athv}{\Av^{\theta}}
\newcommand{\Az}{{A^{z}}}
\newcommand{\bv}{{\bm b}}
\newcommand{\Bv}{{\bm B}}
\newcommand{\Bz}{{B_z}}
\newcommand{\Bc}{B_{\rm c}}
\newcommand{\Bsv}{{\bm B}_{\rm s}}
\newcommand{\Bvs}{{\bm B}_{\rm s}}
\newcommand{\BRv}{{\bm B}_{\rm R}}
\newcommand{\Bveff}{{\bm B}_{\rm eff}}
\newcommand{\Bve}{{\bm B}_{\rm e}}
\newcommand{\betaso}{{\beta_{\rm so}}}
\newcommand{\betasf}{{\beta_\spinflip}}
\newcommand{\betana}{{\beta_{\rm na}}}
\newcommand{\betaw}{{\beta_{\rm w}}}
\newcommand{\cbar}{\avbar{c}}
\newcommand{\cdag}{{c^{\dagger}}}
\newcommand{\chiz}{\chi^{(0)}}
\newcommand{\chio}{\chi^{(1)}}
\newcommand{\chitilo}{\tilde{\chi}^{(1)}}
\newcommand{\chitilz}{\tilde{\chi}^{(0)}}
\newcommand{\chiuni}{\chi_{0}}
\newcommand{\cH}{c_{H}}
\newcommand{\cHdag}{c_{H}^{\dagger}}
\newcommand{\ckv}{c_{\kv}}
\newcommand{\ckvs}{c_{\kv\sigma}}
\newcommand{\ccv}{{\bm c}}
\newcommand{\Cbeta}{C_\beta}
\newcommand{\Cr}{C_\rightarrow}
\newcommand{\Cl}{C_\leftarrow}
\newcommand{\Ci}{C_{\rm i}}
\newcommand{\Ct}{C_t}
\newcommand{\cv}{{\bm c}}
\newcommand{\Cv}{{\bm C}}
\newcommand{\ctil}{\tilde{c}}
\newcommand{\dbar}{\avbar{d}}
\newcommand{\deltaS}{\delta S}
\newcommand{\Deltasd}{\Delta_{sd}}
\newcommand{\dels}{{s_0}}
\newcommand{\ddagg}{d^{\dagger}}
\newcommand{\dtil}{\tilde{d}}
\newcommand{\dv}{{\bm d}}
\newcommand{\dw}{{\rm w}}
\newcommand{\dx}{{d^3 x}}
\newcommand{\Deltatil}{\tilde{\Delta}}
\newcommand{\Dcal}{{\cal D}}
\newcommand{\DOS}{N}
\newcommand{\dos}{\nu}
\newcommand{\doss}{\nu_0}
\newcommand{\dosu}{\dos_{+}}
\newcommand{\dosd}{\dos_{-}}
\newcommand{\dosp}{\dos_{+}}
\newcommand{\dosm}{\dos_{-}}
\newcommand{\DOSV}{{N(0)}}
\newcommand{\DOSom}{{\DOS_\omega}}
\newcommand{\Dv}{{\bm D}}
\newcommand{\ef}{{\epsilon_{\rm F}}}
\newcommand{\eF}{{\epsilon_{\rm F}}}
\newcommand{\eft}{{\epsilon_F \tau}}
\newcommand{\eftoh}{\frac{\epsilon_F \tau}{\hbar}}
\newcommand{\eftauinv}{\frac{\hbar}{\epsilon_F \tau}}
\newcommand{\ekv}{\epsilon_{\kv}}
\newcommand{\ekvs}{\epsilon_{\kv\sigma}}
\newcommand{\ekvp}{\epsilon_{\kv'}}
\newcommand{\ekvps}{\epsilon_{\kv'\sigma}}
\newcommand{\elld}{{\ell_{\rm D}}}
\newcommand{\ells}{\ell_{\rm s}}
\newcommand{\Ev}{{\bm E}}
\newcommand{\Evs}{{\bm E}_{\rm s}}
\newcommand{\Esv}{{\bm E}_{\rm s}}
\newcommand{\ERv}{{\bm E}_{\rm R}}
\newcommand{\ev}{{\bm e}}
\newcommand{\evth}{{\bm e}_{\theta}}
\newcommand{\evph}{{\bm e}_{\phi}}
\newcommand{\evs}{{\bm n}}
\newcommand{\evsz}{{\evs}_0}
\newcommand{\evsph}{(\evph\times\evz)}
\newcommand{\evx}{{\bm e}_{x}}
\newcommand{\evy}{{\bm e}_{y}}
\newcommand{\evz}{{\bm e}_{z}}
\newcommand{\ez}{{\epsilon}_0}
\newcommand{\fl}{{\eta}}
\newcommand{\fltil}{\tilde{\eta}}
\newcommand{\flitil}{\tilde{\fl_{\rm I}}}
\newcommand{\flrtil}{\tilde{\fl_{\rm R}}}
\newcommand{\fli}{{\fl_{\rm I}}}
\newcommand{\flr}{{\fl_{\rm R}}}
\newcommand{\fkv}{f_{\kv}}
\newcommand{\fkvs}{f_{\kv\sigma}}
\newcommand{\fo}{{f(\omega)}}
\newcommand{\fpo}{{f'(\omega)}}
\newcommand{\fomega}{{\omega}}
\newcommand{\fbeta}{f^{\beta}}
\newcommand{\fpin}{{f_{\rm pin}}}
\newcommand{\Fpin}{{F_{\rm pin}}}
\newcommand{\fe}{{f_{\rm e}}}
\newcommand{\fna}{{f_{\rm ref}}}
\newcommand{\Fe}{F}
\newcommand{\Fev}{\Fv}
\newcommand{\Fbeta}{F^{\beta}}
\newcommand{\Fbetav}{\Fv^{\beta}}
\newcommand{\Fbetafactor}{\mu}
\newcommand{\Fhall}{F^{\rm Hall}}
\newcommand{\Fhallv}{\Fv^{\rm Hall}}
\newcommand{\Fren}{F^{\rm ren}}
\newcommand{\Frenv}{\Fv^{\rm ren}}
\newcommand{\Fvna}{\Fv^{\rm ref}}
\newcommand{\Fnav}{\Fv^{\rm ref}}
\newcommand{\Fna}{F^{\rm ref}}
\newcommand{\Fad}{\Fhall}
\newcommand{\Fzad}{F^{\rm (0)ad}}
\newcommand{\Fv}{{\bm F}}
\newcommand{\Fo}{F^{(1)}}
\newcommand{\Fw}{F_{\rm w}}
\newcommand{\Fz}{F^{(0)}}
\newcommand{\Fzv}{\Fv^{(0)}}
\newcommand{\Fdelta}{\delta F}
\newcommand{\Fdeltav}{\delta \Fv}
\newcommand{\Fdel}{\delta \Fo}
\newcommand{\Fdelv}{\delta \Fv^{(1)}}
\newcommand{\Gv}{{\bm G}}
\newcommand{\gv}{{\bm g}}
\newcommand{\gr}{g^{\rm r}}
\newcommand{\gto}{g^{\rm t}}
\newcommand{\ga}{g^{\rm a}}
\newcommand{\Gr}{G^{\rm r}}
\newcommand{\Ga}{G^{\rm a}}
\newcommand{\Gto}{G^{\rm t}}
\newcommand{\Gat}{G^{\overline{\rm t}}}
\newcommand{\Gless}{G^{<}}
\newcommand{\Ggrt}{G^{>}}
\newcommand{\gless}{g^{<}}
\newcommand{\ggrt}{g^{>}}
\newcommand{\Gtil}{\tilde{G}}
\newcommand{\Gcal}{{\cal G}}
\newcommand{\gap}{\Delta_{\rm sw}}
\newcommand{\gammap}{\gamma_{+}}
\newcommand{\gammam}{\gamma_{-}}
\newcommand{\gammav}{\bm{\gamma}}
\newcommand{\gammaz}{\gamma_{0}}
\newcommand{\grst}{|0\rangle}
\newcommand{\grsthc}{\langle0|}
\newcommand{\grft}{|\ \rangle}
\newcommand{\gyro}{\gamma}
\newcommand{\gyroz}{\gyro_{0}}
\newcommand{\hf}{\frac{1}{2}}
\newcommand{\HA}{{H_{A}}}
\newcommand{\HB}{H_{B}}
\newcommand{\Hv}{\bm{H}}
\newcommand{\He}{H_{\rm e}}
\newcommand{\Heff}{H_{\rm eff}}
\newcommand{\Hem}{H_{\rm em}}
\newcommand{\Hex}{H_{\rm ex}}
\newcommand{\Himp}{H_{\rm imp}}
\newcommand{\Hint}{H_{\rm int}}
\newcommand{\HR}{{H_{\rm R}}}
\newcommand{\Hs}{{H_{\rm S}}}
\newcommand{\Hee}{{H_{\rm ee}}}
\newcommand{\Hsf}{{H_\spinflip}}
\newcommand{\Hsr}{{H_\spinflip}}
\newcommand{\Hso}{{H_{\rm so}}}
\newcommand{\Hsd}{H_{sd}}
\newcommand{\Hst}{{H_{\rm ST}}}
\newcommand{\Hw}{H_{\dw}}
\newcommand{\Hz}{H_{0}}
\newcommand{\hbarinv}{\frac{1}{\hbar}}
\renewcommand{\Im}{{\rm Im}}
\newcommand{\ime}{\gamma}
\newcommand{\intinf}{\int_{-\infty}^{\infty}}
\newcommand{\intek}{\int_{-\ef}^{\infty}\! d\epsilon}
\newcommand{\intom}{\int\! \frac{d\omega}{2\pi}}
\newcommand{\intx}{\int\! {d^3x}}
\newcommand{\intk}{\int\! \frac{d^3k}{(2\pi)^3}}
\newcommand{\intr}{\int\! {d^3r}}
\newcommand{\intt}{\int_{-\infty}^{\infty}\! {dt}}
\newcommand{\ioh}{\frac{i}{\hbar}}
\newcommand{\iv}{\bm{i}}
\newcommand{\Ibar}{\overline{I}}
\newcommand{\Iv}{\bm{I}}
\newcommand{\Jex}{{J_{\rm ex}}}
\newcommand{\Jsd}{J_{sd}}
\newcommand{\Js}{{J_{\rm s}}}
\newcommand{\Jv}{\bm{J}}
\newcommand{\js}{j_{\rm s}}
\newcommand{\jsc}{{j_{\rm s}^{\rm c}}}
\newcommand{\jsv}{\bm{j}_{\rm s}}
\newcommand{\Jsv}{\bm{J}_{\rm S}}
\newcommand{\JSv}{\bm{J}_{\rm S}}
\newcommand{\JS}{J_{\rm S}}
\newcommand{\JStotv}{\bm{J}_{S,{\rm tot}}}
\newcommand{\jc}{j_{\rm c}}
\newcommand{\jcv}{\jv_{\rm c}}
\newcommand{\jci}{{{j}_{\rm c}^{\rm i}}}
\newcommand{\jce}{{{j}_{\rm c}^{\rm e}}}
\newcommand{\jatil}{{\tilde{j}_{\rm a}}}
\newcommand{\jctil}{{\tilde{j}_{\rm c}}}
\newcommand{\jcitil}{{\tilde{j}_{\rm c}^{\rm i}}}
\newcommand{\jcetil}{{\tilde{j}_{\rm c}^{\rm e}}}
\newcommand{\jstil}{{\tilde{j}_{\rm s}}}
\newcommand{\jtil}{{\tilde{j}}}
\newcommand{\jv}{\bm{j}}
\newcommand{\kB}{{k_B}}
\newcommand{\kb}{{k_B}}
\newcommand{\kv}{{\bm k}}
\newcommand{\kvxv}{\kv\cdot\xv}
\newcommand{\kvo}{{\kv_1}}
\newcommand{\kvp}{{\kv}'}
\newcommand{\kpq}{{k+\frac{q}{2}}}
\newcommand{\kmq}{{k-\frac{q}{2}}}
\newcommand{\kvpq}{{\kv}+\frac{\qv}{2}}
\newcommand{\kvmq}{{\kv}-\frac{\qv}{2}}
\newcommand{\kvopq}{{\kvo}+\frac{\qv}{2}}
\newcommand{\kvomq}{{\kvo}-\frac{\qv}{2}}
\newcommand{\kvppq}{{\kvp}+\frac{\qv}{2}}
\newcommand{\kvpmq}{{\kvp}-\frac{\qv}{2}}
\newcommand{\kf}{{k_{\rm F}}}
\newcommand{\kF}{{k_{\rm F}}}
\newcommand{\kfpm}{{k_{F\pm}}}
\newcommand{\kfmp}{{k_{F\mp}}}
\newcommand{\kfp}{k_{F+}}
\newcommand{\kfm}{k_{F-}}
\newcommand{\kfu}{k_{F+}}
\newcommand{\kfd}{k_{F-}}
\newcommand{\kfs}{k_{F\sigma}}
\newcommand{\kom}{{k_\omega}}
\newcommand{\Kp}{{K_\perp}}
\newcommand{\ktil}{\tilde{k}}
\newcommand{\lams}{{\lambda_{\rm s}}}
\newcommand{\lamv}{{\lambda_{\rm v}}}
\newcommand{\lamso}{{\lambda_{\rm so}}}
\newcommand{\lambdav}{{\bm \lambda}}
\newcommand{\lamz}{{\lambda_{0}}}
\newcommand{\Le}{{L_{\rm e}}}
\newcommand{\Lez}{{L_{\rm e}^0}}
\newcommand{\Leff}{L_{\rm eff}}
\newcommand{\Lb}{L_{\rm B}}
\newcommand{\Ldw}{L_{\dw}}
\newcommand{\Lsd}{{L_{sd}}}
\newcommand{\Ls}{{L_{\rm S}}}
\newcommand{\Lsw}{L_{\rm sw}}
\newcommand{\Lswdw}{L_{\rm sw-dw}}
\newcommand{\Linv}{{\frac{1}{L}}}
\newcommand{\lstil}{\tilde{l_\sigma}}
\newcommand{\Lv}{{\bm L}}
\newcommand{\mv}{{\bm m}}
\newcommand{\Mv}{{\bm M}}
\newcommand{\Mphi}{{M_{\phi}}}
\newcommand{\Mw}{{M_{\dw}}}
\newcommand{\Ms}{M_{\rm s}}
\newcommand{\MR}{{\rm MR}}
\newcommand{\Mz}{M_{0}}
\newcommand{\mus}{{g\mu_{B}}}
\newcommand{\mub}{\mu_B}
\newcommand{\muB}{\mu_B}
\newcommand{\muz}{\mu_0}
\newcommand{\muu}{\mu_+}
\newcommand{\mud}{\mu_-}
\newcommand{\muspin}{\mu_{\rm s}}
\newcommand{\nel}{n_{\rm e}}
\newcommand{\Ne}{N_{\rm e}}
\newcommand{\nv}{{\bm n}}
\newcommand{\Nv}{{\bm N}}
\newcommand{\Nimp}{N_{\rm i}}
\newcommand{\nimp}{n_{\rm i}}
\newcommand{\Ninv}{\frac{1}{N}}
\newcommand{\Nw}{N_{\dw}}
\newcommand{\nablav}{\bm{\nabla}}
\newcommand{\nvortex}{n_{\rm v}}
\newcommand{\nvz}{{\nv}_0}
\newcommand{\nso}{n_{\so}}
\newcommand{\np}{{n}_+}
\newcommand{\nm}{{n}_-}
\newcommand{\nz}{{n}_0}
\newcommand{\om}{{\omega}}
\newcommand{\omegap}{\omega'}
\newcommand{\Omegatil}{\tilde{\Omega}}
\newcommand{\Omegap}{\Omega'}
\newcommand{\Omegapin}{\Omega_{\rm pin}}
\newcommand{\ompOm}{{\omega+\frac{\Omega}{2}}}
\newcommand{\ommOm}{{\omega-\frac{\Omega}{2}}}
\newcommand{\Omz}{\Omega_0}
\newcommand{\ompOmz}{{\omega+\frac{\Omz}{2}}}
\newcommand{\ommOmz}{{\omega-\frac{\Omz}{2}}}
\newcommand{\Omhf}{\frac{\Omega}{2}}
\newcommand{\phiz}{{\phi_0}}
\newcommand{\Phiv}{\bm{\Phi}}
\newcommand{\PhiB}{{\Phi_{\rm B}}}
\newcommand{\Ptil}{{\tilde{P}}}
\newcommand{\pv}{{\bm p}}
\newcommand{\Pv}{{\bm P}}
\newcommand{\pvh}{\hat{\bm p}}
\newcommand{\PDOS}{P_{\DOS}}
\newcommand{\qv}{{\bm q}}
\newcommand{\qvxv}{\qv\cdot\xv}
\newcommand{\qtil}{{\tilde{q}}}
\newcommand{\ra}{\rightarrow}
\renewcommand{\Re}{{\rm Re}}
\newcommand{\rhow}{{\rho_{\dw}}}
\newcommand{\rhos}{{\rho_{\rm s}}}
\newcommand{\rhoS}{{\rho_{\rm s}}}
\newcommand{\rhoxy}{{\rho_{xy}}}
\newcommand{\RS}{{R_{\rm S}}}
\newcommand{\Rw}{{R_{\dw}}}
\newcommand{\rv}{{\bm r}}
\newcommand{\Rv}{{\bm R}}
\newcommand{\sd}{$s$-$d$}
\newcommand{\sigmav}{{\bm \sigma}}
\newcommand{\sigmaB}{\sigma_{\rm B}}
\newcommand{\sigmab}{\sigma_{\rm B}}
\newcommand{\sigmaz}{\sigma_0}
\newcommand{\sigmas}{\sigma_{\rm s}}
\newcommand{\s}{{\rm s}}
\newcommand{\se}{{s}}
\newcommand{\sev}{{\bm \se}}
\newcommand{\sevsf}{{\bm \se}_\spinflip}
\newcommand{\SE}{\Sigma}
\newcommand{\SEr}{\Sigma^{\rm r}}
\newcommand{\SEa}{\Sigma^{\rm a}}
\newcommand{\SEless}{\Sigma^{<}}
\newcommand{\sgn}{{\rm sgn}}
\newcommand{\sz}{{s}_0}
\newcommand{\sv}{{{\bm s}}}
\newcommand{\seth}{{\se}_\theta}
\newcommand{\seph}{{\se}_\phi}
\newcommand{\sez}{{\se}_z}
\newcommand{\so}{{\rm so}}
\newcommand{\spol}{{M}}
\newcommand{\spinflip}{{\rm sr}}
\newcommand{\svtil}{\tilde{\bm s}}
\newcommand{\stil}{\tilde{\se}}
\newcommand{\stilz}{\stil_{z}}
\newcommand{\stilpm}{\stil^{\pm}}
\newcommand{\stilpmz}{\stil^{\pm(0)}}
\newcommand{\stilpma}{\stil^{\pm(1{\rm a})}}
\newcommand{\stilpmb}{\stil^{\pm(1{\rm b})}}
\newcommand{\stilpmo}{\stil^{\pm(1)}}
\newcommand{\stilpara}{\stil_{\parallel}}
\newcommand{\stilperp}{\stil_{\perp}}
\newcommand{\Simpv}{{{\bm S}_{\rm imp}}}
\newcommand{\Simp}{{S_{\rm imp}}}
\newcommand{\Stot}{{S_{\rm tot}}}
\newcommand{\Stotv}{\bm{S}_{\rm tot}}
\newcommand{\Sh}{{\hat {S}}}
\newcommand{\Svh}{{\hat {\Sv}}}
\newcommand{\Sv}{{{\bm S}}}
\newcommand{\Svz}{{{\bm S}_0}}
\newcommand{\sumx}{{\int\! \frac{d^3x}{a^3}}}
\newcommand{\sumk}{{\sum_{k}}}
\newcommand{\sumkv}{{\sum_{\kv}}}
\newcommand{\sumqv}{{\sum_{\qv}}}
\newcommand{\sumom}{\int\!\frac{d\omega}{2\pi}}
\newcommand{\sumOm}{\int\!\frac{d\Omega}{2\pi}}
\newcommand{\sumomOm}{\int\!\frac{d\omega}{2\pi}\int\!\frac{d\Omega}{2\pi}}
\newcommand{\thickness}{{d}}
\newcommand{\thetaz}{{\theta_0}}
\newcommand{\tr}{{\rm tr}}
\newcommand{\To}{{\rm T}}
\newcommand{\Ta}{\overline{\rm T}}
\newcommand{\Tc}{{\rm T}_{C}}
\newcommand{\Tct}{{\rm T}_{\Ct}}
\newcommand{\tcmp}{\tau}
\newcommand{\tcmpi}{\tau_{\rm I}}
\newcommand{\tcmpinf}{\tau_{\infty}}
\newcommand{\tcmpz}{\tau_{0}}
\newcommand{\tcmpzp}{\tau_{0}'}
\newcommand{\Torqv}{{\bm \tau}}
\newcommand{\torque}{{\tau}}
\newcommand{\torquev}{{\bm \torque}}
\newcommand{\Torquev}{{\bm \tau}}
\newcommand{\torqueve}{\torquev}
\newcommand{\torquee}{\torque}
\newcommand{\torquew}{{\torque_{\dw}}}
\newcommand{\tautil}{{\tilde{\tau}}}
\newcommand{\taup}{\tau_{+}}
\newcommand{\taum}{\tau_{-}}
\newcommand{\tauw}{\tau_{\dw}}
\newcommand{\taus}{\tau_{\rm s}}
\newcommand{\tauso}{\tau_\so}
\newcommand{\tausf}{\tau_\spinflip}
\newcommand{\betasr}{{\beta_{\rm sr}}}
\newcommand{\thetast}{\theta_{\rm st}}
\newcommand{\ttil}{{\tilde{t}}}
\newcommand{\tinf}{t_\infty}
\newcommand{\tz}{t_0}
\newcommand{\Ubar}{\overline{U}}
\newcommand{\Ueff}{U_{\rm eff}}
\newcommand{\Uz}{U_0}
\newcommand{\Uv}{U_V}
\newcommand{\vc}{{v_{\rm c}}}
\newcommand{\ve}{{v_{\rm e}}}
\newcommand{\vev}{{\vv_{\rm e}}}
\newcommand{\vv}{\bm{v}}
\newcommand{\vs}{{v_{\rm s}}}
\newcommand{\vsv}{{\vv_{\rm s}}}
\newcommand{\vw}{v_{\rm w}}
\newcommand{\vf}{{v_F}}
\newcommand{\vimp}{v_{\rm i}}
\newcommand{\vi}{{v_{\rm i}}}
\newcommand{\Vi}{{V_{\rm i}}}
\newcommand{\Vso}{v_{\rm so}}
\newcommand{\vtil}{{{v_0}}}
\newcommand{\Vpin}{{V}_{\rm pin}}
\newcommand{\Vinv}{\frac{1}{V}}
\newcommand{\vz}{{v_0}}
\newcommand{\Vz}{{V_0}}
\newcommand{\Vcal}{{\cal V}}
\newcommand{\Vztil}{{\tilde{V_0}}}
\newcommand{\Ws}{{W_{\rm S}}}
\newcommand{\Xtil}{{\tilde{X}}}
\newcommand{\xv}{{\bm x}}
\newcommand{\Xv}{{\bm X}}
\newcommand{\xvp}{{\bm x}_{\perp}}
\newcommand{\xw}{{z}}
\newcommand{\Xz}{{X_0}}
\newcommand{\Zs}{Z_{\rm S}}
\newcommand{\Zz}{Z_{0}}
\newcommand{\ztil}{u}
\newcommand{\zh}{\hat{z}}
\newcommand{\zv}{\bm {z}}

\newcommand{\gkv}{{\bm \gamma}_{\bm k}}
\newcommand{\hgkv}{\hat{{\bm \gamma}}_{\bm k}}
\newcommand{\hav}{\hat{{\bm \alpha}}}
\newcommand{\hkv}{\hat{{\bm k}}}
\newcommand{\gqv}{{\bm \gamma}_{\bm q}}
\newcommand{\tgkv}{\tilde{{\bm \gamma}}_{\bm k}}
\newcommand{\vare}{\varepsilon}
\newcommand{\beqa}{\begin{eqnarray}}
\newcommand{\eeqa}{\end{eqnarray}}
\newcommand{\beq}{\begin{equation}}
\newcommand{\eeq}{\end{equation}}
\newcommand{\bars}{\bar{\sigma}}
\newcommand{\upa}{\uparrow}
\newcommand{\doa}{\downarrow}
\newcommand{\Gt}{\tilde\Gamma_2}
\newcommand{\Gi}{\tilde\Gamma_1}
\newcommand{\bR}{\bar{R}}
\newcommand{\bA}{\bar{A}}
\newcommand{\varex}{\xi}
\newcommand{\zvhat}{\hat{\zv}}
\newcommand{\AR}{{\boldmath {\cal A}}}

\newcommand{\alphatil}{\widetilde{\alpha}}
\newcommand{\alphavhat}{\hat{\alphav}}
\newcommand{\betav}{{\bm \beta}}
\newcommand{\kappaE}{\kappa_{\rm E}}
\newcommand{\kappaIE}{\kappa_{\rm IE}}
\newcommand{\IE}{{\rm IE}}
\newcommand{\mev}{{\mv}}
\newcommand{\omegaF}{\omega_{\rm F}}
\newcommand{\omegaR}{\omega_{\rm R}}
\newcommand{\omegaP}{\omega_{\rm p}}

\title{
Theory of anomalous optical properties of bulk Rashba conductor \\ 
}
\author{Junya Shibata}
\email{j_shibata@toyo.jp}
\affiliation{
Department of Electrical, Electronic and Communications Engineering,  
Toyo University, Kawagoe, Saitama, 350-8585, Japan}
\author{Akihito Takeuchi}
\affiliation{Department of Physics and Mathematics, Aoyama Gakuin University, Sagamihara, Kanagawa 252-5258, Japan}
\author{Hiroshi Kohno}
\affiliation{
Department of Physics, Nagoya University, Furo-cho, Chikusa-ku, Nagoya, 464-8602, Japan}
\author{Gen Tatara}
\email{gen.tatara@riken.jp}
\affiliation{RIKEN Center for Emergent Matter Science (CEMS), 
2-1 Hirosawa, Wako, Saitama, 351-0198 Japan}
\date{\today}

\begin{abstract}
The 
Rashba interaction induced when inversion symmetry is broken in solids is a key interaction connecting spin and charge for realizing novel magnetoelectric cross-correlation effects. 
Here, we theoretically explore the optical properties of a bulk Rashba conductor by calculating the transport coefficients at finite frequencies. 
It is demonstrated that the combination of direct and inverse Edelstein effects leads to a softening of the plasma frequency for the electric field perpendicular to the Rashba field, resulting in a hyperbolic electromagnetic metamaterial.  
In the presence of magnetization, a significant enhancement of anisotropic propagation (directional dichroism) is predicted because of interband transition edge singularity. 
Based on an effective Hamiltonian analysis, the dichroism is demonstrated to be driven by 
 toroidal and quadratic moments of the magnetic Rashba system.
The effective theory of the cross-correlation effects 
has the same mathematical structure as that of insulating multiferroics.
\end{abstract}  
\maketitle

\section{Introduction}
The spin$-$orbit interaction is a relativistic interaction connecting the spin and orbital motion of charged elementary particles. 
This interaction plays essential roles in conversion between electric signals and magnetic ones \cite{Hirsch99,Saitoh06}. 
Of particular interest is the Rashba spin$-$orbit interaction$-$the one that arises when the inversion symmetry of the system is broken \cite{Rashba60}. 
It is represented by the 
quantum mechanical Hamiltonian
\begin{align}
  H=\frac{\hat{\pv}^2}{2m}-\frac{\alphav}{\hbar}\cdot(\hat{\pv}\times\sigmav),
\end{align}
where $m$ is the electron mass, $\hbar$ is the Planck constant divided by $2\pi$, 
$\alphav$ is a vector representing the direction and strength of the Rashba field and 
$\hat{\pv}$ and $\sigmav$ are the momentum operators for the electron and  vector of Pauli matrices, respectively. 
The interaction leads to various cross-correlation effects in electromagnetism, i.e. 
a mixing of electric and magnetic fields. 
A straightforward consequence of the Rashba interaction in the dc limit is the spin polarization of a conduction electron  induced when an electric current is applied (called the Rashba$-$Edelstein effect) \cite{Edelstein90}, and the electric current generated when a magnetic field is applied (the inverse Edelstein effect) \cite{Shen14}.
The Rashba-induced spin polarization has been predicted to be highly useful for magnetization switching by electric current \cite{Obata08,Manchon09}.
Magnetoelectric effects have been intensively studied in insulating materials (multiferroics), where magnetization and electric polarization are mutually coupled \cite{Tokura06}.
A conducting Rashba system is unique in this sense, since magnetoelectric effects couple magnetization and electric current instead of electric polarization; thus, the effect is expected to be smoothly integrated into conventional electronic devices. 

Although the Rashba interaction realized in two-dimensional electron gases 
in semiconductors has generally been weak, 
a strong Rashba interaction of $\alpha=3$ eV\AA\ was recently 
observed at surfaces of noble metals doped with Bi \cite{Ast07}.
Metallic surfaces and interfaces have now become highly important in spintronics for magnetization control and spin-to-charge conversion with high efficiency \cite{Miron10,Sanchez13}. 
It has been observed that 
topological insulators, the extreme cases of Rashba systems where 
the quadratic term in the dispersion disappears, 
have intriguing features such as a coupled collective mode of spin and charge (spin plasmon) on the surface \cite{Raghu10}.
Bulk BiTeI, a semiconductor where carrier electrons are introduced by Bi vacancies \cite{Horak85}, was recently observed to have a large Rashba splitting of 
$\alpha=3.85$ eV\AA \cite{Ishizaka11}; a large 
magneto-optical (Kerr) effect \cite{Demko12} and a giant spin-polarized photocurrent \cite{Ogawa14} were also discovered.

The coupling of electric and magnetic degrees of freedom by the Rashba spin$-$orbit interaction leads to intriguing optical responses, including anisotropic propagation (directional dichroism) and electromagnetic metamaterial behavior;  
metamaterials are materials that exhibit extraordinary electromagnetic 
characteristics such as negative refraction \cite{Smith04}.
Negative refraction was originally discussed by Veselago 
and was considered to occur if the electric permittivity and magnetic permeability were  simultaneously negative \cite{Veselago68}.
It has in fact been achieved  by designing resonators for electric and magnetic responses with the same resonant frequencies  in GHz \cite{Pendry00} and THz \cite{Moser05} ranges. 
For shorter wavelengths, small structures comparable to the wavelength are necessary, 
and thus, artificial structures would not be suitable.  
Instead, natural materials become promising, since their structures 
can be designed down to the atomic size.
In fact, oxide superconductors with layered structures,  such as La$_{2-x}$Sr$_x$CuO$_4$ and Bi$_2$Sr$_2$CaCu$_2$O$_{8+\delta}$, behave as  metallic materials in 
the direction parallel to the layers and as dielectric materials in the direction perpendicular to the layers in the infrared frequency range \cite{Narimanov15}. 
Such materials with strong anisotropic electric properties (called hyperbolic materials) 
exhibit negative refraction when an interface parallel to the metallic plane is subjected to an 
electromagnetic wave. These materials are also useful for filters and for enabling hyperlenses to go beyond the diffraction limit \cite{Poddubny13,Narimanov15}.
Tetradymites, 
such as Bi$_2$Te$_3$ and Bi$_2$Se$_3$, 
have been observed to be a new class of natural materials exhibiting a a hyperbolic nature for visible light \cite{Esslinger14}. 

In this paper, we present a quantitative and consistent theoretical analysis of the optical properties of a bulk Rashba conductor by evaluating the correlation functions from a microscopic standpoint. 
First, we shall demonstrate that bulk Rashba systems such as BiTeI indeed exhibit naturally significant hyperbolic metamaterial properties in the infrared regime because of spin$-$charge mixing. 
These unique behaviors of Rashba systems are not observed in topological insulators, 
for which the bare plasma frequency is not defined by the quadratic dispersion.
Second,  we demonstrate that in the presence of magnetization or a magnetic field, 
the system exhibits directional dichroism because of breaking of both spatial inversion and time reversal symmetries. 
The effect is strongly enhanced by a singularity at the interband transition edges, 
the stronger of which is predicted to occur in the infrared and visible light regime for BiTeI.
Our analysis establishes  a bridge from the theoretical side between the optical properties discussed mostly on phenomenological grounds and solid-state transport phenomena 
based on the microscopic quantum mechanical viewpoint. 
The qualitative equivalence of the dc transport and optical responses 
is noted experimentally as well as from the symmetry viewpoints \cite{Rikken01,Rikken05}.
The equivalence is natural from our analysis, since both phenomena are governed by a same conductivity tensor evaluated at different frequencies.

%

\section{Circular dichroism due to spin$-$charge conversion}

When an electric field is applied, the Rashba interaction generates electron spin polarization (the so-called Edelstein effect) \cite{Edelstein90} (Fig.  1 a) 
whose density reads 
\begin{align}
\sev_{\rm E}=\kappaE\alphavhat\times\Ev, \label{sE}
\end{align}
where $\alphavhat\equiv\frac{\alphav}{\alpha}$ is a unit vector along the Rashba field and $\kappaE$ is a coefficient.
The induced magnetization is $\Mv_{\rm E}\equiv -\hbar\gamma \sev_{\rm E}$, 
where $\gamma$ is the gyromagnetic ratio.
The interaction also induces the inverse effect, where an electric current density is induced by an applied magnetic field $\Bv$, namely
\begin{align}
\jv_{\IE}=\kappaIE\alphavhat\times\Bv,\label{jIE}
\end{align}
where $\kappaIE$ is a coefficient.

These cross-correlation effects also affect the optical properties.
The inverse Edelstein current (\ref{jIE}) is written by using Faraday's law as (in the momentum and angular frequency representation)
$
\jv_{\IE}= \frac{\kappaIE}{\omega}[\alphavhat\times(\qv\times\Ev)]$. 
The Edelstein effect (\ref{sE}) also leads to a magnetization current density,
$
\jv_{\rm E}\equiv \nabla\times \Mv_{\rm E}$. 
The coefficients of the Edelstein effect and its inverse effect are related as 
\begin{align}
  \kappaIE= i\hbar\gamma \omega \kappaE, 
\end{align}
since both the spin density (\ref{sE}) and current density (\ref{jIE}) are written in terms of the uniform component of the spin$-$current correlation function $\chi_{sj}^{ij}$ (the superscripts $ij$ denote the direction).
We thus see that the sum of the two generated currents is  
$
\jv_{\IE}+ \jv_{\rm E} = i\hbar\gamma {\kappaE}[\qv(\alphavhat\cdot\Ev)-\alphavhat(\qv\cdot\Ev)] 
$ and that the conductivity tensor has off-diagonal elements linear in $\qv$.
The off-diagonal elements result in a rotation of the incident electric field 
{\color{black} around} a rotation axis 
$\betav\equiv \hbar\gamma \kappaE(\qv\times\alphavhat)$ 
{\color{black} by an} angle proportional to {\color{black}  $|\betav |$.} 
The dispersion reads  $q^2=\frac{\omega^2}{c^2}\pm i\muz\beta\omega$ 
($\muz$ is the magnetic permeability) for a circular polarization $\pm$, and hence, the light velocity depends on both the incident direction 
(sign of $\qv\times\alphavhat$) and the circular polarization 
(directional circular dichroism). 
This result is natural from a symmetry viewpoint, since breaking of inversion symmetry alone leads to the so-called natural optical activity, which affects only circularly polarized light \cite{Wagniere99}.

\section{Hyperbolic metamaterial resulting from direct and inverse Edelstein effects}

Peculiar behaviors of unpolarized light arise if the diagonal components of the conductivity tensor are modified by the Rashba interaction that occurs when the direct and inverse Edelstein effects are  combined. 
In fact, the magnetization $\Mv_{\rm E}$ induced by the Edelstein effect generates an electric current due to the inverse Edelstein effect: 
$\jv_{{\rm IE}\cdot{\rm E}} \equiv \kappaIE (\alphavhat\times\Mv_{\rm E})= -\hbar\gamma\kappaE\kappaIE [\alphavhat\times(\alphavhat\times\Ev)]$.
The resultant polarization ${\bm P} = {\bm j}_{\rm IE\cdot E}/(-i\omega)$ is opposite to the applied electric field when $\Ev\perp\alphavhat$; 
in other words, the mixing of spin and charge excitations reduces the plasma frequency 
(Fig. 1 a).  
Microscopically, this softening effect is represented in the conductivity tensor 
in the long-wavelength ($q\ra0$) limit: 
\begin{align}
\sigma_{ij}(\qv=0, \omega)
= \frac{ie^2}{\omega+i\eta}\frac{n_{\rm e}}{m}\biggl(
\delta_{ij}(1 + C(\omega)) -\hat\alpha_i \hat\alpha_j C(\omega)
\biggr),
\label{conductivitytensor}
\end{align}
where
\begin{align}
C(\omega)\equiv -\frac{4\tilde{\alpha}^2}{n_{\rm e}}\epsilon_{\rm F}\int\frac{d^3k}{(2\pi)^3} \frac{\gamma_k s_k}{(\hbar\omega+i\eta)^2-4\gamma_k^2}%
\end{align}
is a function representing the magnitudes of the Rashba$-$Edelstein and inverse Edelstein effects (plotted in Fig. 1 b). 
Here, 
$s_\kv\equiv\sum_{\sigma=\pm}\sigma f_{\kv\sigma}$ is the electron spin polarization ($\sigma=\pm$ denotes spin), $\gamma_\kv\equiv |\gammav_\kv|$,  $\gammav_\kv\equiv \kv\times\alphav$,  $f_{\kv\sigma}\equiv [e^{\ekvs/(\kb T)}+1]^{-1}$ is the Fermi distribution function, ($\kb$ and $T$ are the Boltzmann constant and temperature, respectively), 
$\ekvs\equiv \frac{\hbar^2k^2}{2m}-\sigma\gamma_\kv-\ef$ ($\ef$ and $\kf$ are the Fermi energy and Fermi wave vector, respectively) and $\eta$ is a damping parameter.

In terms of $C(\omega)$, the Rashba$-$Edelstein coefficient is
$\kappaE=-\frac{e\hbar\nel}{m\alpha}\Im \frac{C(\omega)}{\omega}$ ($\Im$ denotes the imaginary part).
The function $\Im C(\omega)$ is finite for a broad frequency range lower than the higher interband transition edge $\omega_+$, where $\omega_\pm\equiv 4\omegaF  \alphatil(\sqrt{1+\alphatil^2}\pm\alphatil)$  are the transition edges,  with 
$\alphatil\equiv \frac{m\alpha}{\hbar^2\kf}$ being a dimensionless Rashba strength parameter and $\omegaF\equiv \frac{\eF}{\hbar}$ (see Supplementary Information).
For BiTeI, $\omegaF$, the plasma frequency $\omegaP$, and $\omega_-$ are in the infrared regime, whereas $\omega_+$ is in the visible frequency.

Choosing $\alphav $ along the $z$-axis, the dielectric tensor 
$
\vare_{ij}(\qv=0,\omega) = \delta_{ij} + \frac{1}{\vare_{0}}\frac{i}{\omega}
\sigma_{ij}(\qv=0,\omega) 
$
reads 
\begin{align}
\vare_{zz}
 &= 1-\frac{\omegaP^2}{\omega(\omega+i\eta)}\equiv \vare_{z}
\nnr
\vare_{xx}=\vare_{yy}
&=1-\frac{\omegaP^2}{\omega(\omega + i\eta)}
(1+ C(\omega))\equiv \vare_{x},
\label{e-perp}
\end{align}
where $\omegaP = \sqrt{{e^2n_{\rm e}}/{\vare_{0}m}}$ 
is the bare plasma frequency ($\nel$ is the electron density).
As observed in equation (\ref{e-perp}), 
the plasma frequency in the $xy$ plane is indeed reduced by the Rashba interaction to 
$
  \omegaR \equiv \omegaP\sqrt{1+\Re C(\omegaR)}
$. 

The frequency range between the bare plasma frequency $\omegaP$ and 
the reduced plasma frequency $\omegaR$ is of particular interest, 
since the system is insulating in the direction of the Rashba field but 
metallic in the perpendicular direction (Fig. 2 a,b). 
Materials with such anisotropy are called hyperbolic because of their hyperbolic dispersion \cite{Narimanov15}.
In fact, choosing the wavevector $\qv$ in the $xz$ plane, 
the dispersion relation is  $c^2q^2-\omega^2\vare_{x}=0$ for $E_{y}$ 
($E_i$ denotes the $i$-component of the electric field) and  
\begin{align}
\frac{q^{2}_{z}}{\vare_{x}} + \frac{q^2_{x}}{\vare_{z}} = \frac{\omega^2}{c^2} 
\label{disp1}
\end{align}
for $E_{x}$ and $E_{z}$.
When both $\Re\ \vare_x$ and $\Re\ \vare_z$ are positive, 
the dispersion is elliptic; however, it is hyperbolic if  one of these values is negative 
(Fig. 2 c,d). 
Note that 
in equation (\ref{disp1}) 
the component $q_{z}$ is governed by $\vare_{x}$ with positive real part, 
whereas $q_{x}$ is by $\vare_{z}$ with negative real part; thus, 
the wave propagation 
in the $z(x)$-direction is insulating (metallic).

Hyperbolic materials exhibit the following peculiar optical properties. 
When the interface is parallel to the metallic plane, the incident light tends to be refracted in  the negative direction, since the metallic nature reflects the light 
(Fig.  2 a). 
This behavior is observed in the dispersion curve (Fig. 2 c), 
which reveals that the $x$-component of the group velocity is negative. 
The Poynting vector $\Sv$ also has a negative $x$-component. 
Notably, negatively refracted light is focused along the insulating direction, 
as observed in the plot of the transmittance defined by the ratio of the flow of $\Sv$ as 
$T\equiv \frac{\Sv\cdot\zvhat}{\Sv_{\rm in}\cdot\zvhat}$ 
($\Sv_{\rm in}$ is the incident Poynting vector) (Fig.  2 e,g). 
In fact, the angle of refraction, determined by $\tan \theta_S\equiv \frac{\Sv\cdot\hat{\xv}}{\Sv\cdot\hat{\zv}}$, does not exceed a threshold value $\theta_S^{\rm max}=\tan^{-1}\lt(\frac{\Re(1/\vare_z)}{\Re(q_z/\vare_x)}\frac{\omega}{c}\rt)$ and the absolute value of the threshold angle is small, 
since no solution exists for small  $|\Re \ q_z|$ (Fig. 2 c). 
For $\omega/\omegaP=0.78$, $\theta_S^{\rm max}=-13^\circ$ 
(Fig. 2 g), 
the incident light even for a large angle of incidence is concentrated in a direction within $13^\circ$ from the direction perpendicular to the interface.
This focusing effect is stronger for smaller frequencies close to $\omegaR$, 
whereas the distribution is broader for larger frequencies  (Fig. 2 c,g,i). 

In contrast, when the interface is perpendicular to the metallic plane 
(Fig. 2 b), 
the light is mostly reflected when the angle of incidence is small 
(Fig. 2 f), 
since  no solution exists for small $|q_z|$ (Fig. 2 d). 
For transmitted light, in contrast to the parallel case, 
the angle of refraction cannot be small, and thus, 
the refracted waves propagate close to the interface (Fig. 2 h,j).
An interesting feature of the refracted wave is the backward wave, 
where the wave vector $\qv$ and the Poynting vector $\Sv$ are in opposite directions 
when the propagation is perpendicular to the interface 
(Fig. 2 b,d)
\cite{Poddubny13}. 

For BiTeI, the maximum plasma frequency known so far is $\omegaP=2.5\times10^{14}$ Hz (corresponding to a wavelength of $7.5\mu$m) for 
$\nel=8\times 10^{25}$ m$^{-3}$ and $\ef=0.2$ eV \cite{Demko12}.
Using $\alpha=3.85$ eV\AA and $\kf\sim 0.1$ \AA$^{-1}$, we have $\alphatil\sim 1$, and thus, $C(\omega)\sim -0.4$ for frequencies $\omega\lesssim\omegaP$ (Fig. 1 b), 
resulting in $\omegaR/\omegaP=0.77$ ($\omegaR=1.9\times 10^{14}$ Hz, corresponding to the wavelength of $9.8\mu$m). 
Thus, hyperbolic behavior arises in the infrared regime. 

For experimental realization,  the penetration depth $\lambda$  of the system, defined by $\lambda\equiv \frac{1}{\Im [q_z]}$, is also essential. 
The imaginary part of the relative permittivity is on the order of $0.1$ in the hyperbolic regime, and the imaginary part of $q_z$ calculated from  equation (\ref{disp1}) is 
$\Im [q_z]\simeq \frac{\omega}{c}\Im[\vare]$. 
For ${\omega}/\omegaP\sim1$, we have 
$\lambda \simeq 12\mu$m, and the absorption of the light is neglected if 
the sample thickness is smaller.

\section{Directional dichroism of the magnetic Rashba system}

So far, we have considered a pure Rashba conductor where time reversal symmetry is maintained. 
We now discuss the case without time reversal symmetry, as realized in a thin film of a Rashba conductor attached to a ferromagnet with uniform magnetization or in the presence of an external magnetic field.
We thus consider the $sd$ exchange interaction  
\begin{align}
   H_{sd} &= \Jsd\Mv\cdot \sigmav ,
\end{align}
where $\Mv$ is a dimensionless magnetization vector and $\Jsd$ represents the strength of the exchange interaction .
 
The conductivity tensor including the effect of $\Mv$ to the linear order is (the coefficients are defined in the Supplementary Information)
\begin{align}
  \sigma_{ij}^M
  =&  -\frac{e^2 n_{\rm e}\Jsd}{m} \biggl[ 
   \frac{\hbar }{\ef^2} D(\omega)
   (\hat{\bm \alpha}\cdot\Mv)\biggl[
    \vare_{ijk}\hat{\alpha}_{k} -\frac{i}{2\kf\alphatil}(\hat{\alpha}_i\epsilon_{jkl}+\hat{\alpha}_j\epsilon_{ikl} )\hat{\alpha}_k q_l \biggr]
\nnr
& -\frac{1}{\ef\kf\omega} \biggl(  E_{1}(\omega)
[(\hat{\bm \alpha}\times\Mv ) \cdot \qv] \left[\delta_{ij}- \hat\alpha_{i}\hat\alpha_{j}\right]
 +\frac{E_{2}(\omega)}{2}\left[
 (\hat{\bm \alpha}\times \Mv)_{i}q_{\perp,j} 
  + (\hat{\bm \alpha}\times\Mv)_{j}q_{\perp,i}
\right]  \nnr
& - \frac{E_{3}(\omega)}{2}\left[ M^{\perp}_{i}(\hat{\bm \alpha}\times \qv)_{j} + (\hat{\bm \alpha}\times\qv)_{i}M^{\perp}_{j} \right] 
    \biggr) \biggr]
, \label{sigmaMq}
\end{align}
where $\qv_{\perp}\equiv \qv-\alphavhat(\alphavhat\cdot\qv)$ and $\Mv^{\perp}=\Mv-(\Mv\cdot\hat{\bm \alpha})\hat{\bm \alpha}$.
The first term of equation (\ref{sigmaMq}) 
is proportional to $(\hat{\bm \alpha}\cdot\Mv)$ and corresponds to the anomalous Hall effect induced by the magnetization parallel to the Rashba field, which leads to an enormous magneto-optical response (Kerr effect) \cite{Demko12}.
Other terms having diagonal components linear in the wave vector $\qv$ induce directional dichroism when $\alphavhat\times\Mv$ is finite.
We consider the case where the interface is in the $xy$ plane with the magnetization along the $y$-axis and the light incident to the $xz$ plane.
For the electric field in the $y$-direction, the dispersion relation including the effect of magnetization (equation (\ref{sigmaMq}))  leads to 
\begin{align}
q \simeq \frac{\omega}{c}\sqrt{\vare_{x}}+\frac{1}{2}\frac{\omega_{\rm p}^2}{c^2}\frac{\Jsd M}{k_{\rm F}\vare_{\rm F}}
E_{13} (\hat{\qv}\cdot\AR),
\end{align}
when $\dfrac{\omega_{\rm p}}{c \kf}\dfrac{\Jsd M}{\ef} \ll 1$, 
where $E_{13}\equiv E_1+E_3$, $\hat{\qv}=\qv/q$ 
and $\AR\equiv (\alphavhat\times\hat\Mv)$ ($\hat\Mv\equiv \Mv/|\Mv|$) is a moment that governs the dichroism. 
In the broad frequency range below $\omega_+$, $\Im E_{13}$ is on the order of unity, 
whereas it has singularities at the transition edges ($\omega=\omega_\pm$) 
(Fig. 1 b). 
More precisely, the dominant contribution to the function $\Im E_{13}(\omega)$ behaves  as $\Im \frac{d^2 C}{d \omega^2}$, and, since $\Im C\propto \theta(\omega_\pm-\omega)\sqrt{\omega_\pm-\omega}$ ($\theta(x)$ is a step function) close to the edge at $\omega_\pm$,  
$\Im E_{13}$ diverges at the transition edges if $\eta=0$ as $\Im E_{13} \propto (\omega_\pm-\omega)^{-\frac{3}{2}}$. 
Considering a finite damping, the directional dichroism is strongly enhanced by a factor of
\begin{align}
  \Im E_{13}
  \sim \lt(\frac{\eta}{\hbar\omegaP}\rt)^{-\frac{3}{2}}, 
\end{align}
at the transition edges.
Let us define the {\color{black} non-reciprocity} ratio as the ratio of the transmission rate for the angle $\theta_q$ and $\pi-\theta_q$ as 
$\eta_{\rm D}\equiv \frac{T_{\theta_q}-T_{\pi-\theta_q}}{T_{\theta_q}+T_{\pi-\theta_q}}$.
For the system with thickness $t$, $\eta_{\rm D}=\tanh \frac{t}{\ell}$, 
where $\ell^{-1}\equiv \Im q$.
For $\theta_q=45^\circ$ and $t=10\mu$m, $\eta_{\rm D}\sim 0.25\times 10^{-3}$ in the infrared regime with $\Im E_{13}\sim 1$ for $\Jsd M/\ef=0.1$, which is already large enough for experimental detection. 
At the edge $\omega_\pm$, $\eta_{\rm D}$ is strongly enhanced to be approximately  0.24 if $\eta/(\hbar\omegaP)=0.01$.

Mathematically, the edge singularity results from the dichroism arising from the correlation function linear in $q$, which is obtained by expansion with respect to the angular frequency.
An additional condition for  imposing finite $\AR$ was necessary for the singularity in the Rashba case.The singularity does not exist in the coefficient $D$, and only a cusp structure exists at $\omega_-$ (Fig. 3 b) 
(the \textquoteleft singularity\textquoteright\ in Ref. \cite{Lee11} is a cusp.).
In the presence of of many interband transitions, 
the singularity is still expected to be present, since each transition contributes independently to the singularity.


We have demonstrated that the non-reciprocal optical property is determined by the mutual orientation of $\qv$ and a vector $\AR$. 
This vector is known as the Rashba-induced effective vector potential that drives electron spin flow \cite{Takeuchi12,Kim12,Nakabayashi14}, and in this context, 
the directional dichroism in the magnetic Rashba system is a result of the Doppler shift of the light due to the flow induced by the vector potential \cite{Kawaguchi15}. 
The factor $\qv\cdot\AR$ essential for directional dichroism is consistent with the observation of dc transport \cite{Rikken05}, where the electric resistance in the presence of applied electric and magnetic fields ($\Ev$ and $\Bv$) was observed to be anisotropic depending on $\qv\cdot(\Ev\times\Bv)$ ($\qv$ is the wave vector of electron), as observed by the replacement $\alphav\rightarrow\Ev$ and $\Mv\rightarrow\Bv$.

The effective vector potential $\AR$ is also interpreted as a toroidal moment 
(a vector that couples with the rotation of magnetic field) and has been discussed  intensively in multiferroics \cite{Spaldin08}.
In fact, the isotropic contributions of our dichroism result (\ref{sigmaMq}) 
is described by an effective Hamiltonian 
\begin{align}
  H_{\rm d} = -g \AR\cdot(\Ev\times \Bv)-h Q_{ij}E_i B_j,  \label{Hd}
\end{align}
where $g$ and $h$ are coefficients and $Q_{ij}=\hat{M}_i\hat\alpha_j+\hat{M}_j\hat\alpha_i$ (see Supplementary Information). 
The interaction (\ref{Hd}) is the interaction Hamiltonian proposed to describe cross correlation effects in insulator multiferroics, and the vector $\AR$ and $Q_{ij}$ in this context are  toroidal and  quadrupole moments, respectively \cite{Spaldin08}.
The form of interaction between  the toroidal moment and the electromagnetic fields 
in equation (\ref{Hd}) suggests that
the edge singularity is a common feature of toroidal moment-induced directional dichroism. In fact, the energies of lights having positive and negative $\qv\cdot\AR$ get shifted  according to the interaction;  
this energy difference results in directional dichroism 
if the interband transition matrix element depends on the energy.
As a result, the dichroism occurs proportional to the energy derivative of the interband matrix element, and  the edge singularity arises whenever the matrix element vanishes at the edge $\omega_{\rm e}$ faster than linear, i.e. proportional to 
$(\omega_{\rm e}-\omega)^{\xi}$ with $0<\xi<1$. 

In the real-space representation, the isotropic part of the current responsible for the dichroism (contribution of equation (\ref{sigmaMq})) 
is 
$\jv_{\rm d}=\nabla\times\Mv_{\rm d}+\frac{\partial}{\partial t}\Pv_{\rm d}$, where 
$M_{{\rm d},i} \equiv  -g(\AR\times\Ev)_i-h Q_{ij}E_j$ and 
$P_{{\rm d},i} \equiv g(\AR\times \Bv)_i+h Q_{ij}B_j$ are the effective magnetization and electric polarization induced in metals by the toroidal and quadrupole moments, respectively.
These expressions are the metallic analogues of those in multiferroics, 
and our study indicates that the cross-correlation physics discussed so far for insulators is universal and can be applied to metals without changing their mathematical structure.

\newpage

\newpage
\noindent
{\bf Figure 1}
\begin{figure}[H]\centering
\includegraphics[width=0.6\hsize]{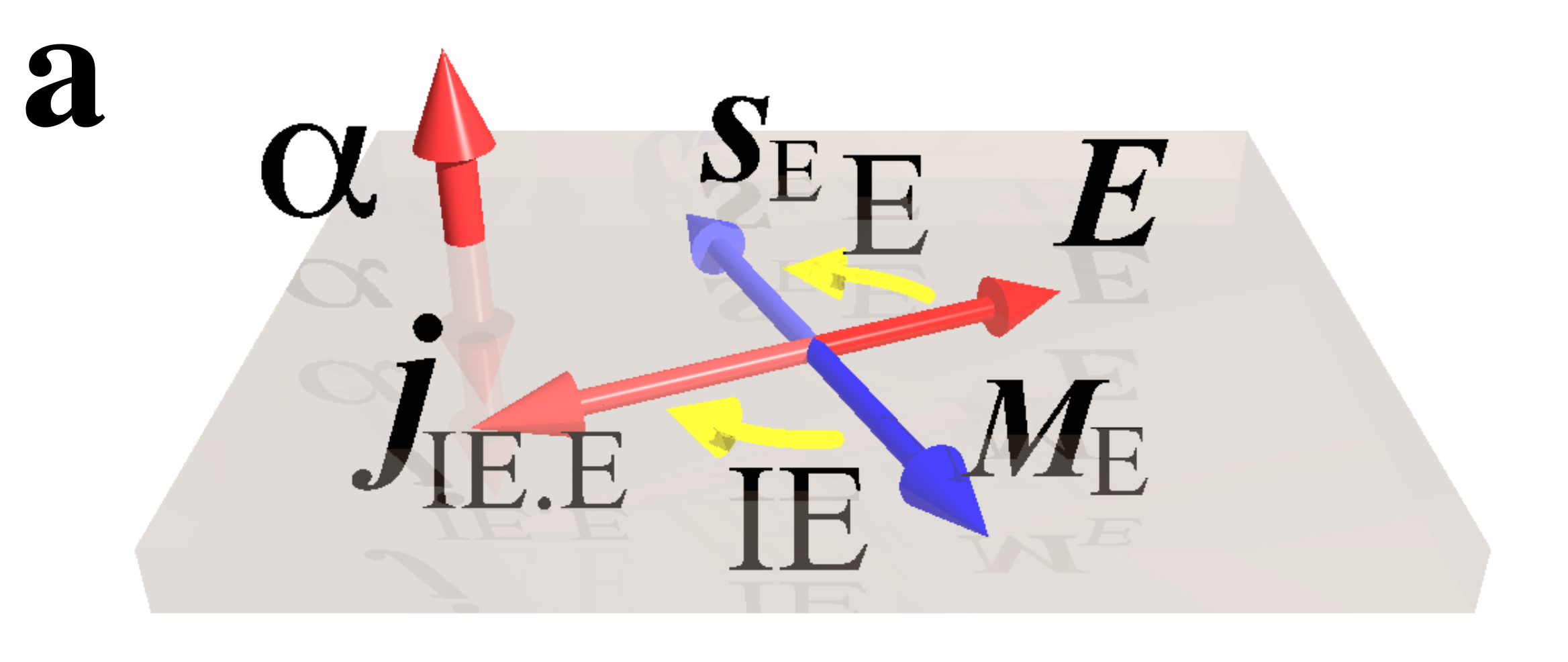}
\includegraphics[width=0.48\hsize]{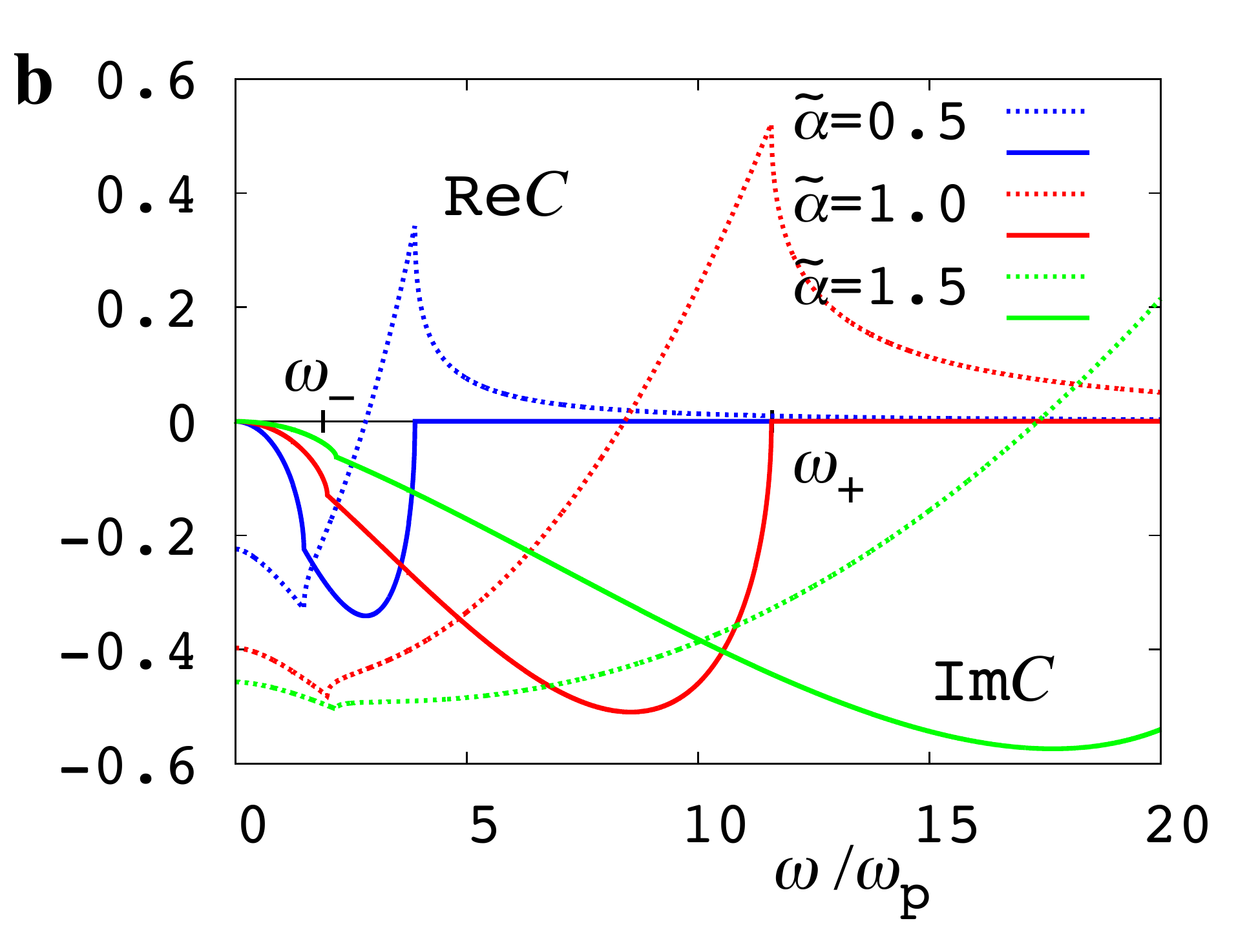}
\includegraphics[width=0.48\hsize]{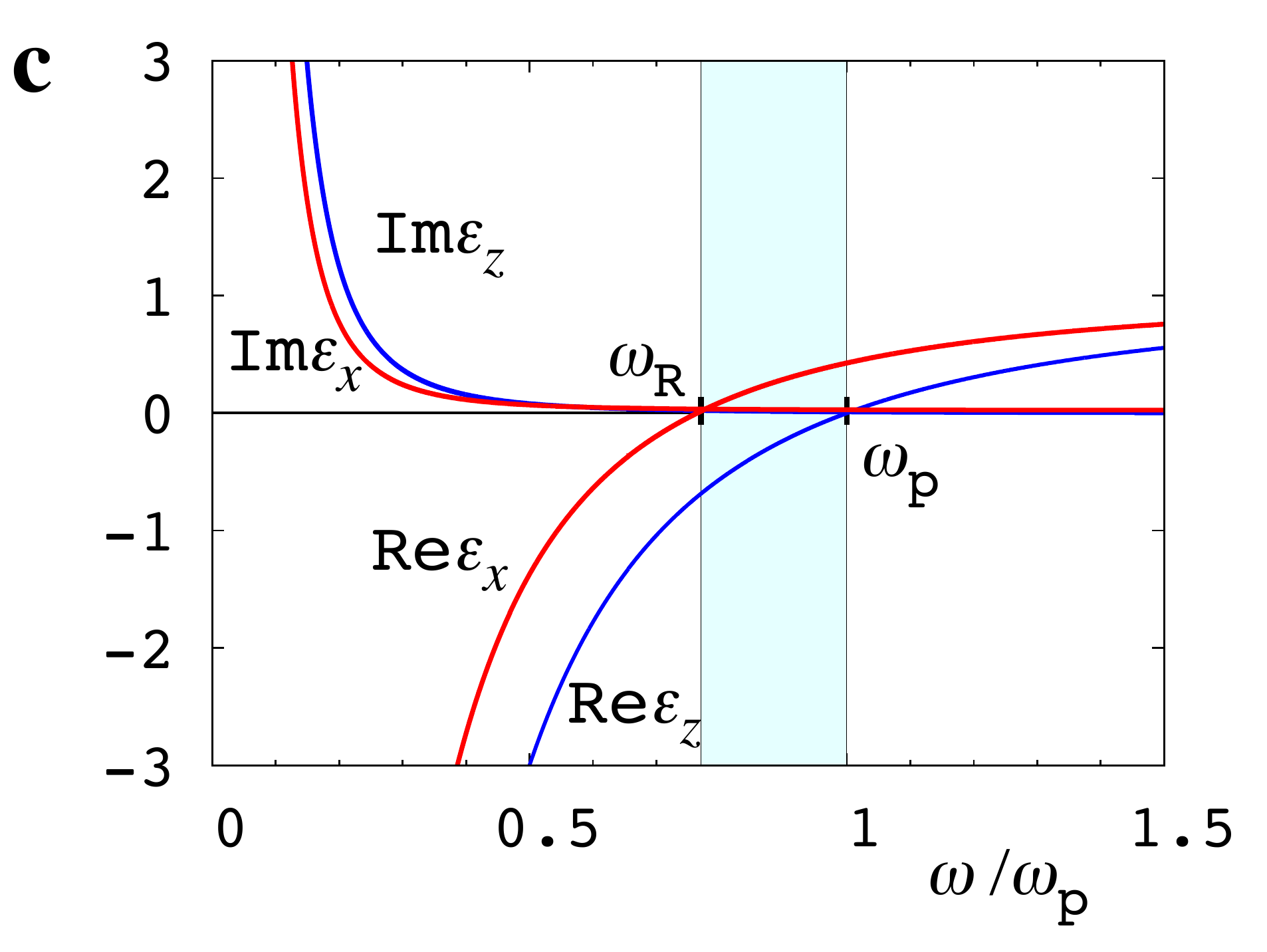}
\end{figure}
\noindent
{\small
Figure 1. 
a: Schematic figure showing the cross-correlation effects in the plane perpendicular to the Rashba field $\alphav$. The Edelstein (E) effect generates spin polarization $\sev_{\rm E}$ from the applied electric field $\Ev$, and the inverse Edelstein (IE) effect generates electric current $\jv_{{\rm IE}\cdot{\rm E}}$ from the induced magnetization $\Mv_{\rm E}$ in the direction  opposite to the applied field. 
\label{FIGEIE}
b: The real (dotted lines) and imaginary (solid lines) parts of the function $C(\omega)$ are plotted as functions of  $\omega/\omegaP$ for $\alphatil=0.5,1,1.5$. 
The interband transition edges $\omega_\pm$ and imaginary part of $E_{13}$ 
governing directional dichroism in the presence of magnetization are shown for the case of $\tilde\alpha=1$. \label{FIGComega}
c: Real and imaginary parts of $\vare_{x}(\omega)$ and 
$\vare_{z}(\omega)$
as a function of  $\omega/\omegaP$ for $\tilde{\alpha}=1.0$ and
$\eta/(\hbar\omegaP) = 0.01$.
The shaded region between $\omegaR$ and $\omegaP$ is the hyperbolic regime.
\label{FIGepsilon}
}


\newpage
\noindent 
{\bf Figure 2}
\begin{figure}[H]\centering
\includegraphics[width=0.41\hsize]{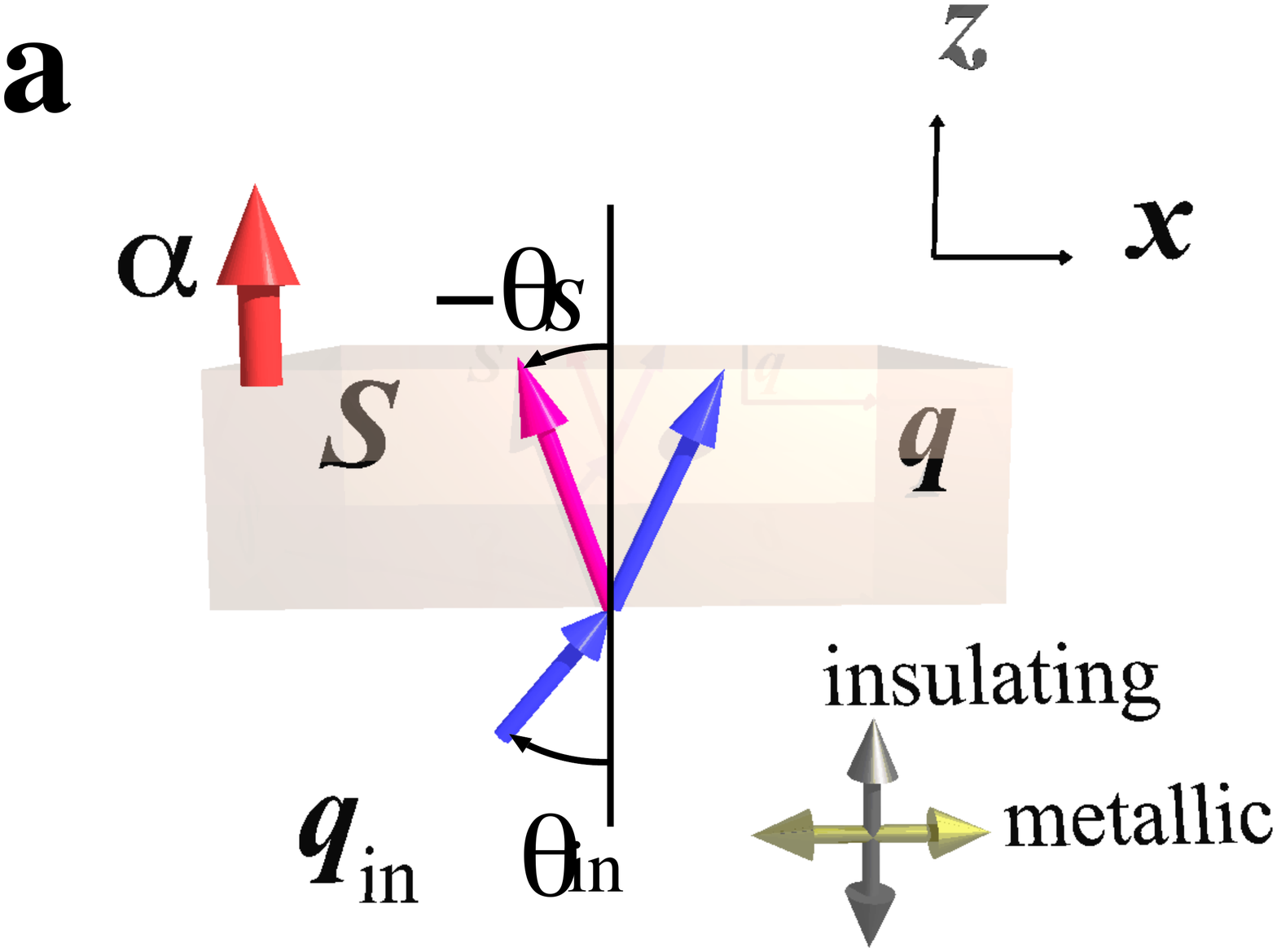}
\includegraphics[width=0.41\hsize]{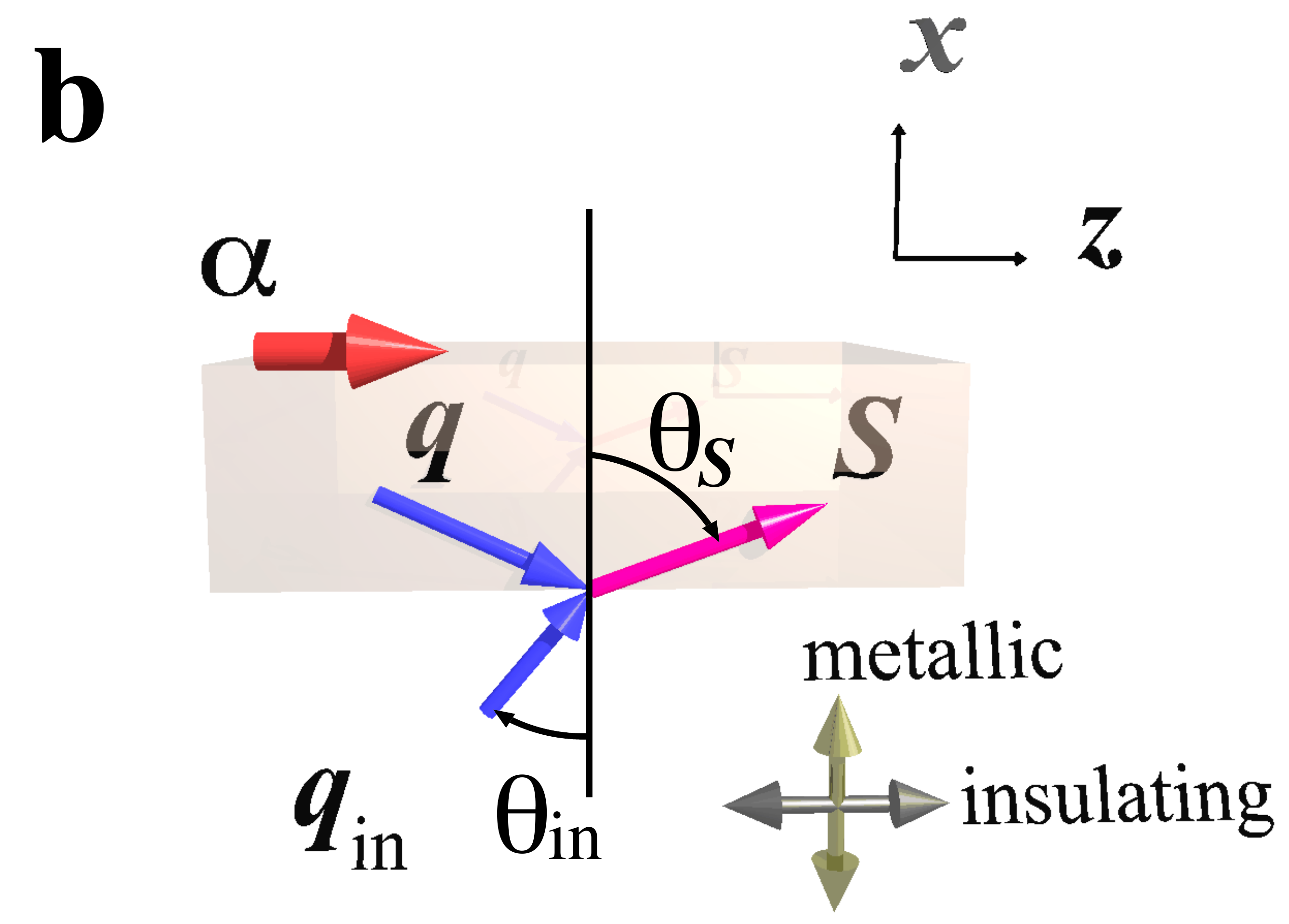}\\
\includegraphics[width=0.41\hsize]{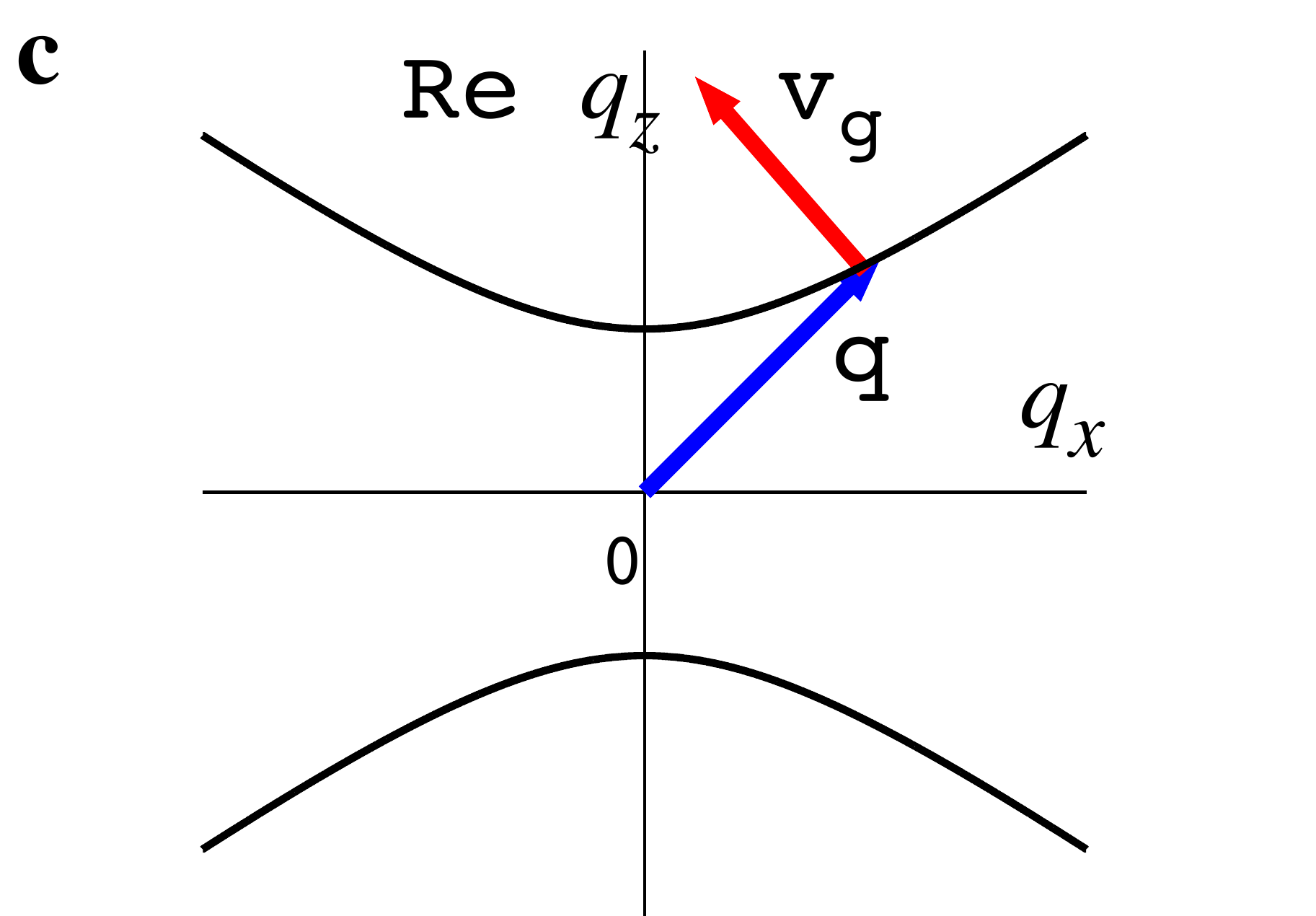}
\includegraphics[width=0.41\hsize]{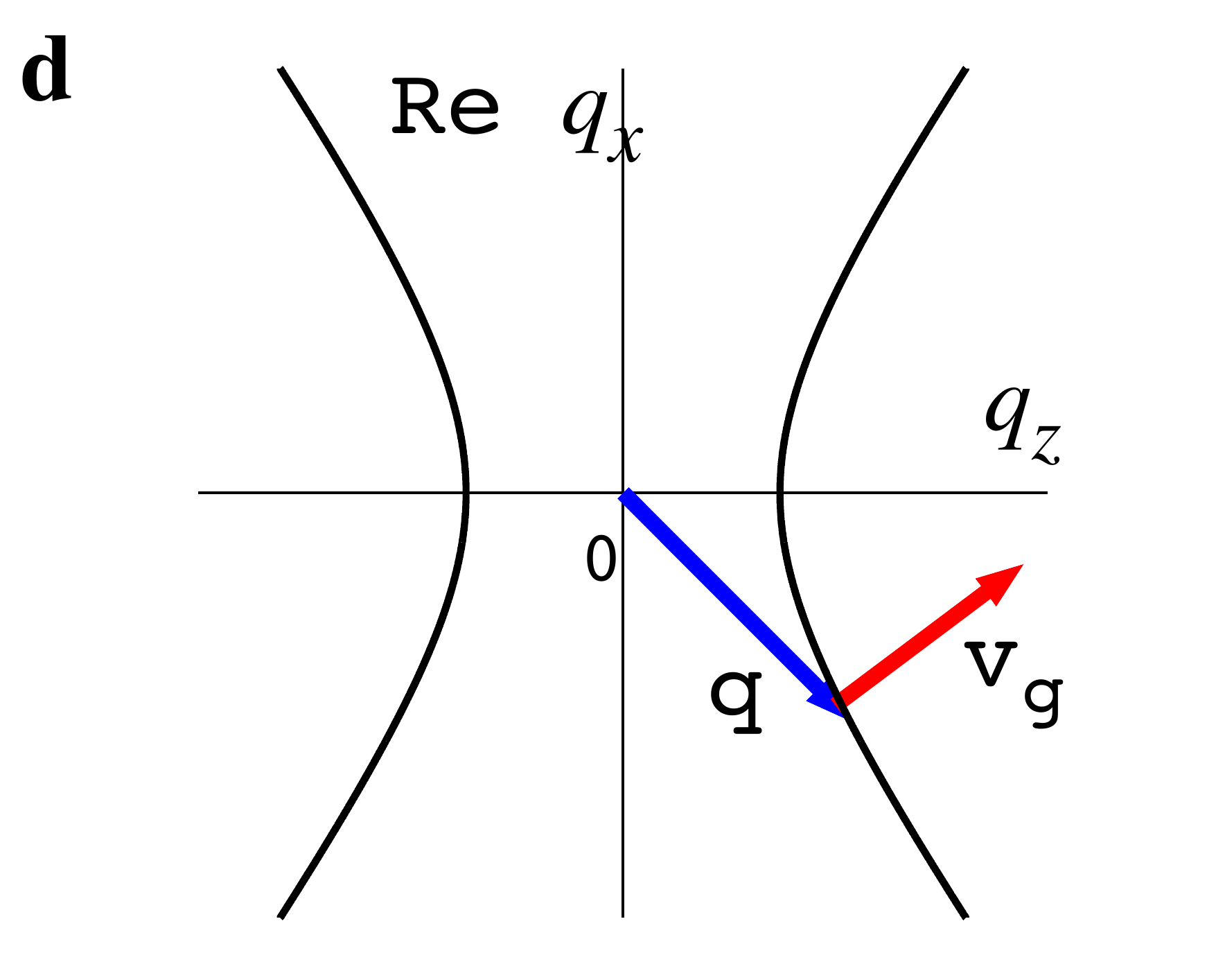}\\
\includegraphics[width=0.45\hsize]{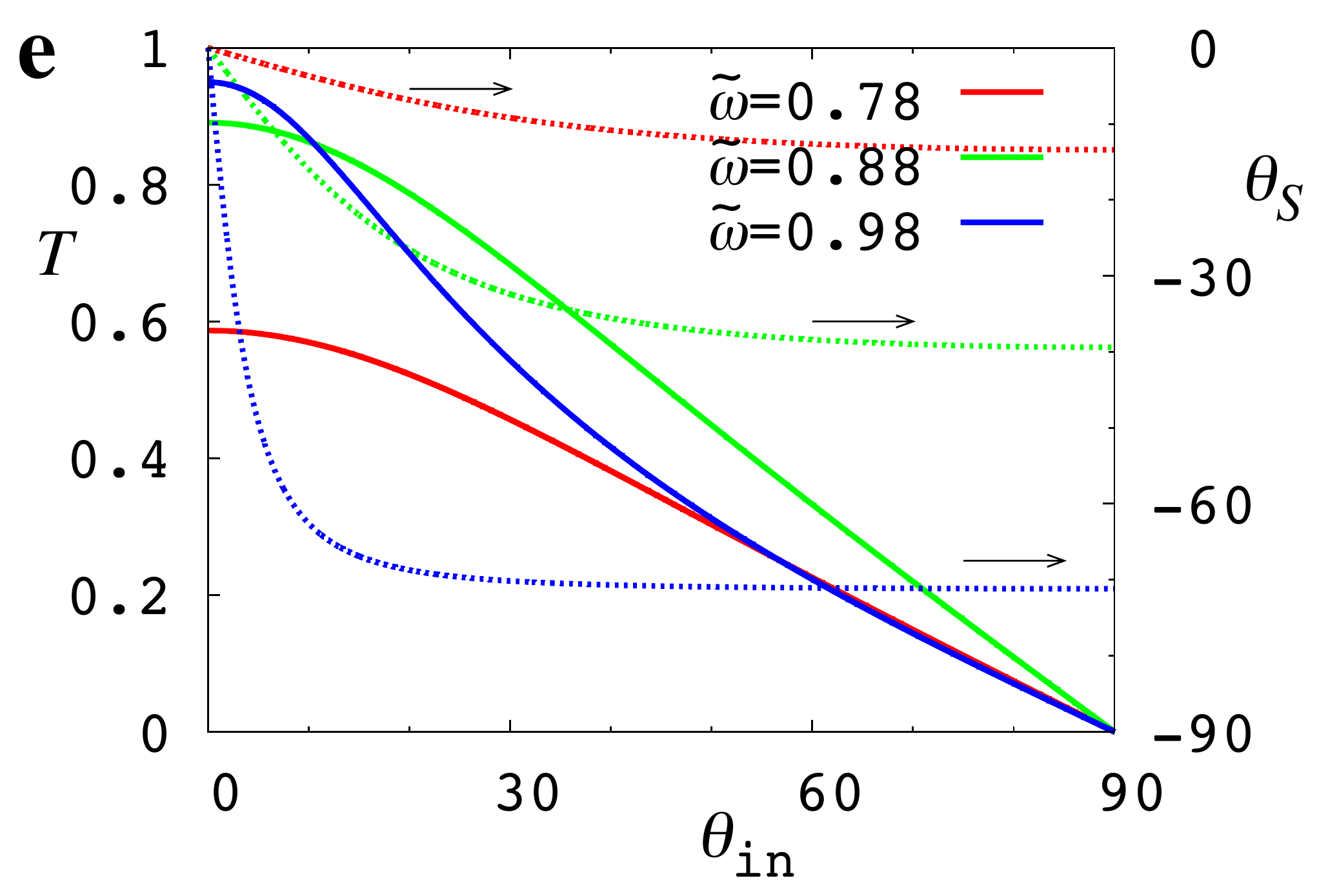}
\includegraphics[width=0.45\hsize]{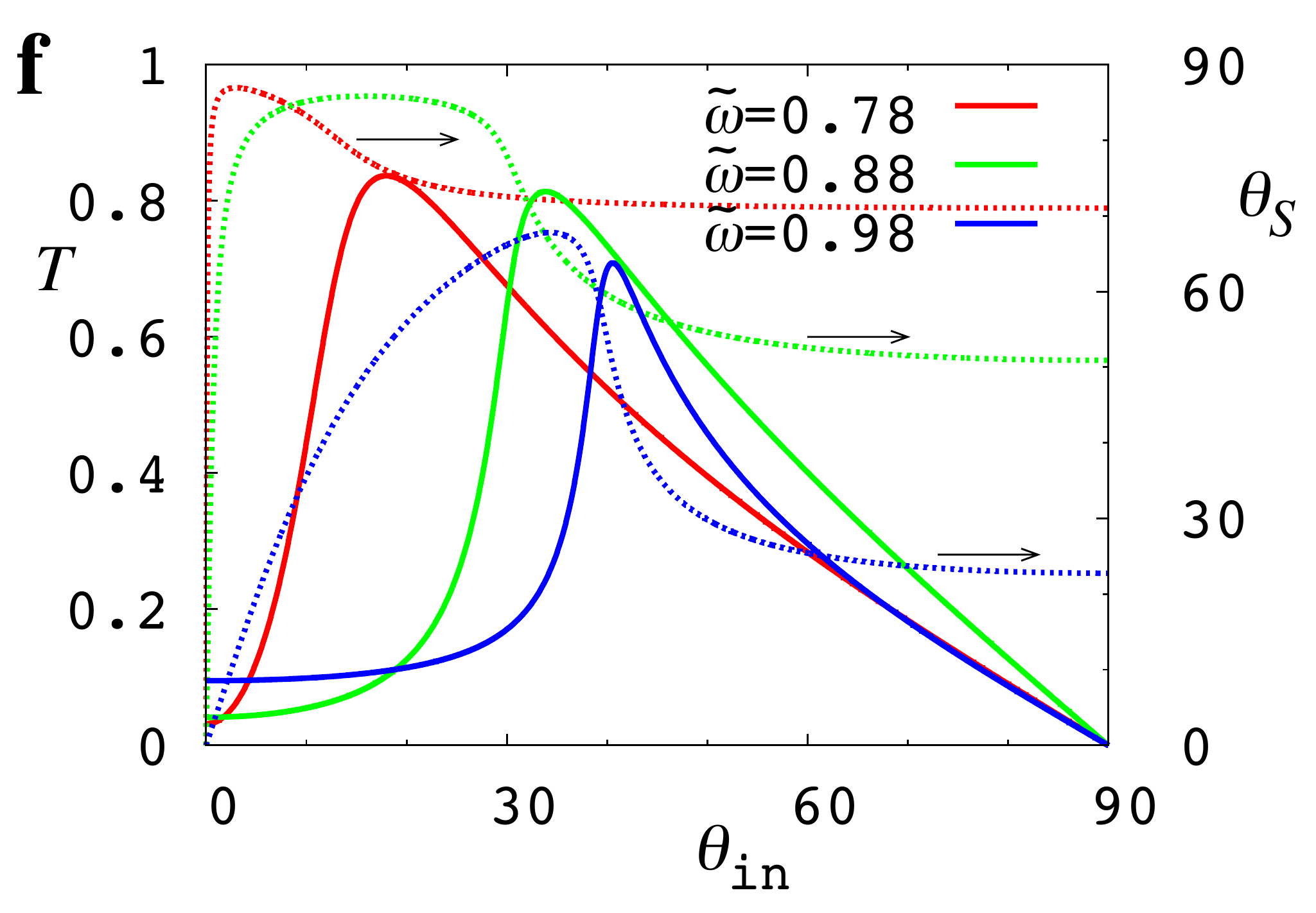}\\
\includegraphics[width=0.45\hsize]{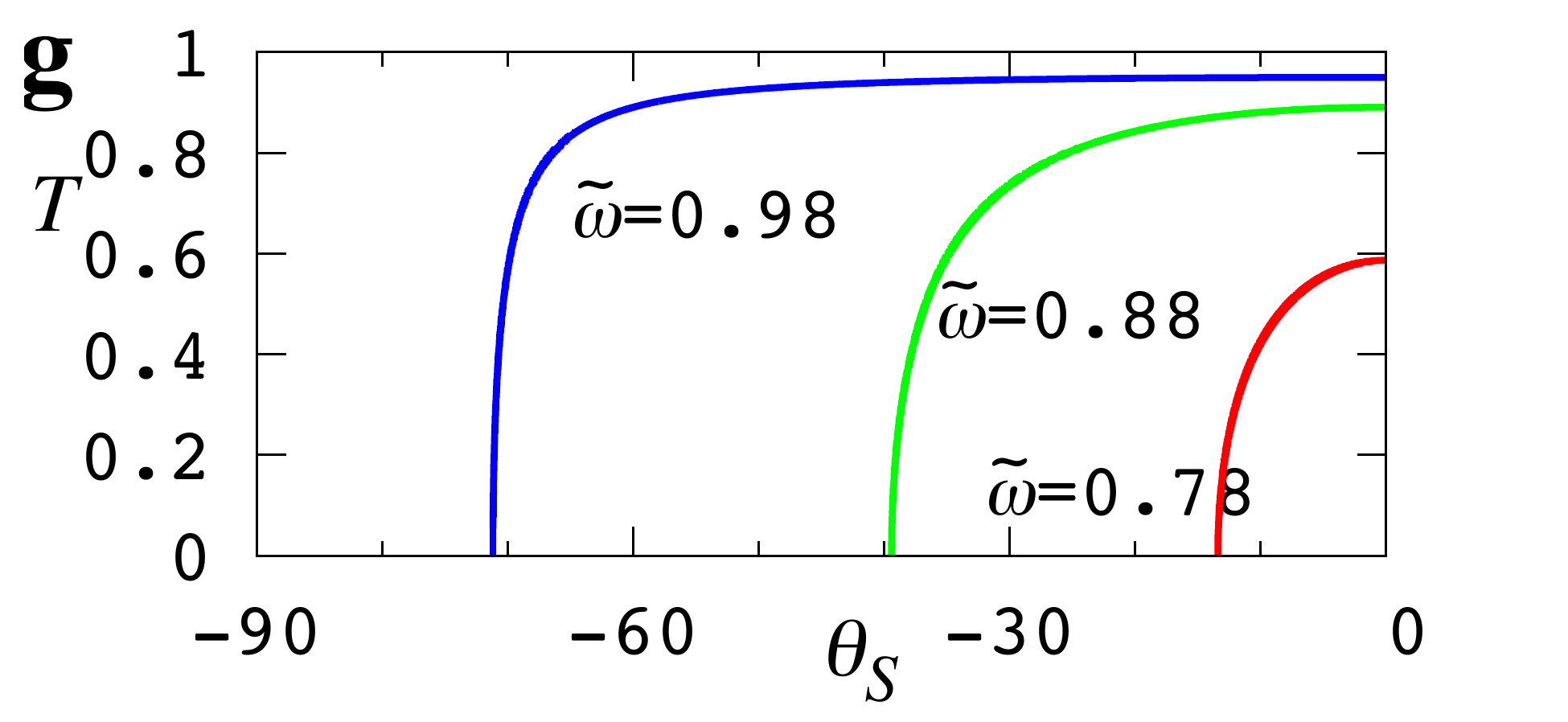}
\includegraphics[width=0.45\hsize]{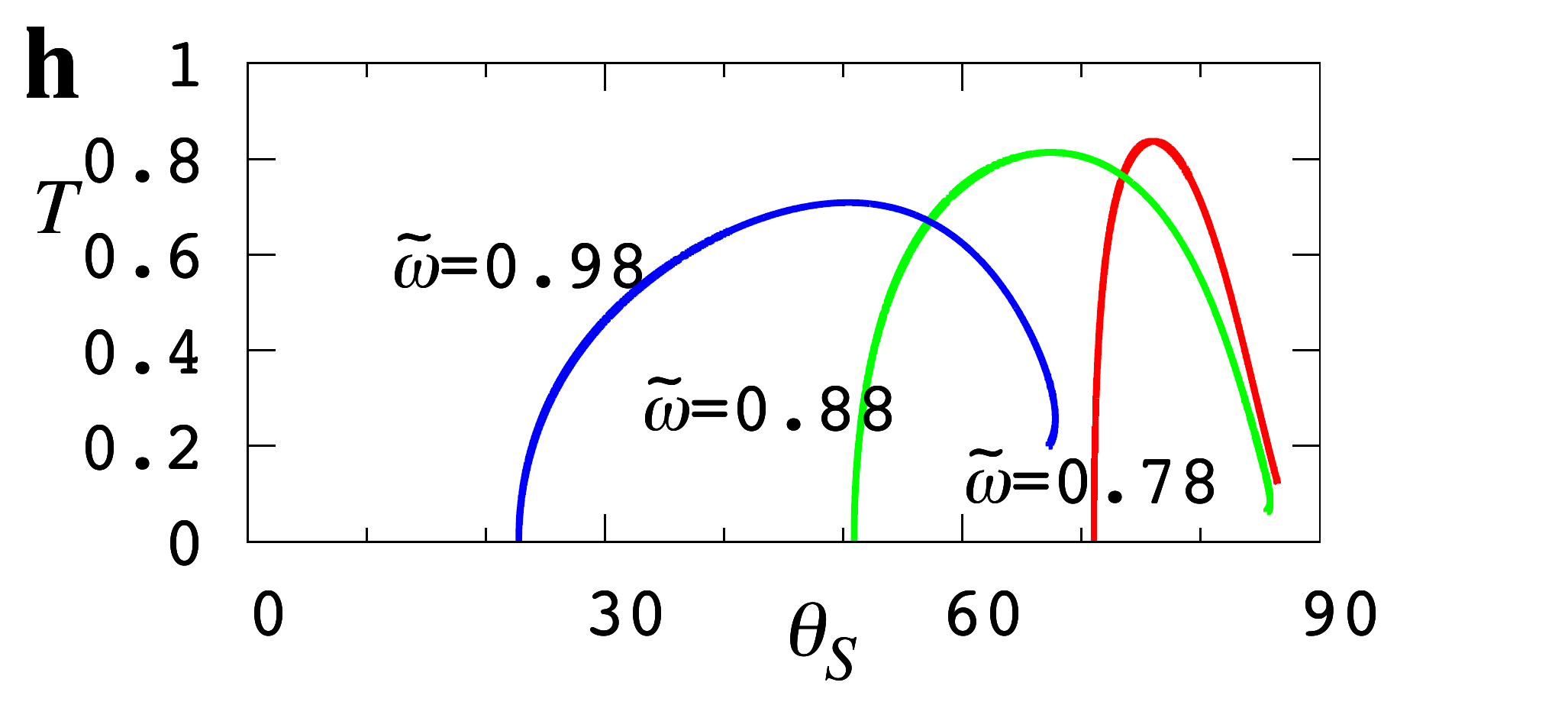}\\
\includegraphics[width=0.21\hsize]{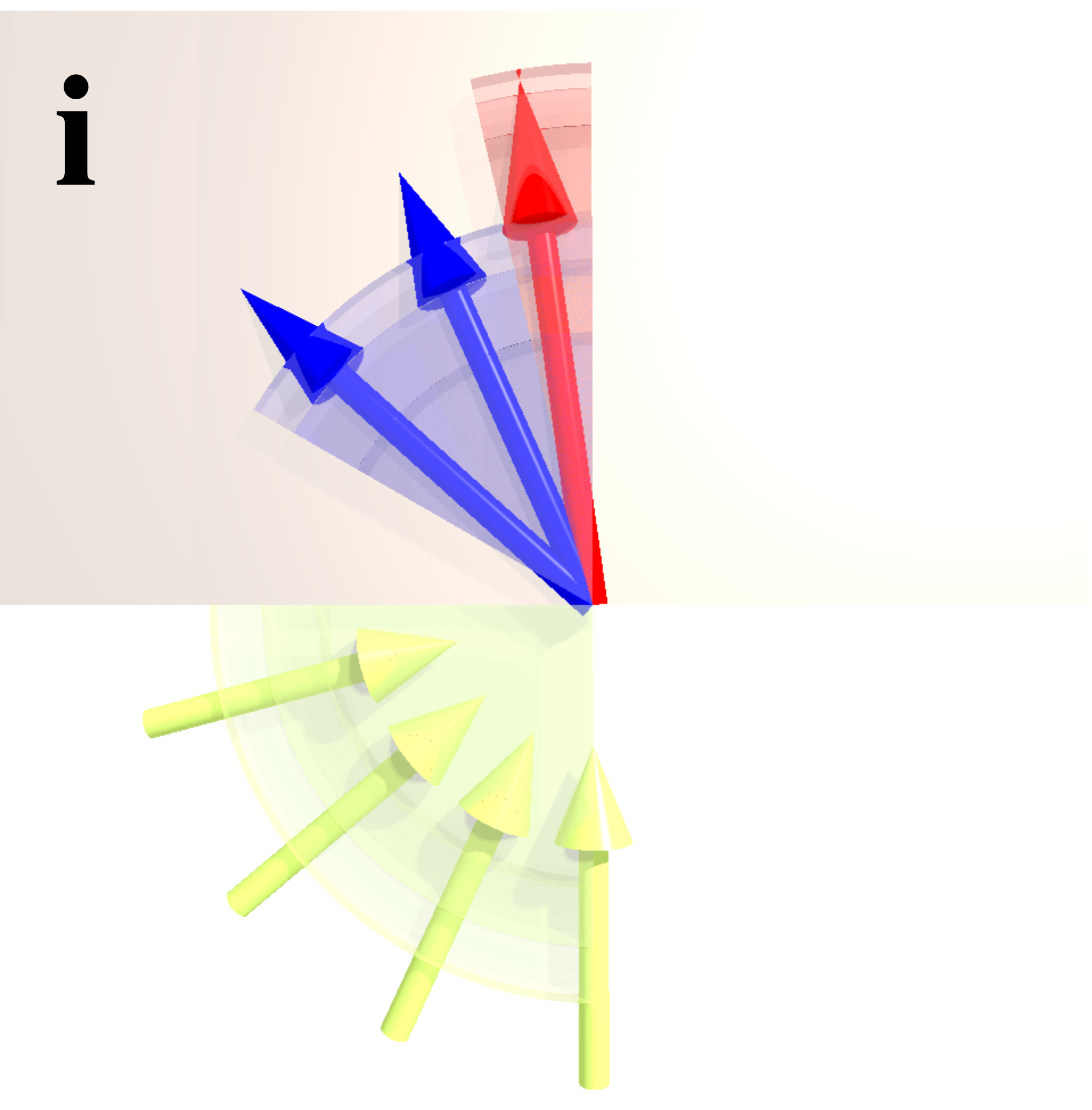}
\includegraphics[width=0.21\hsize]{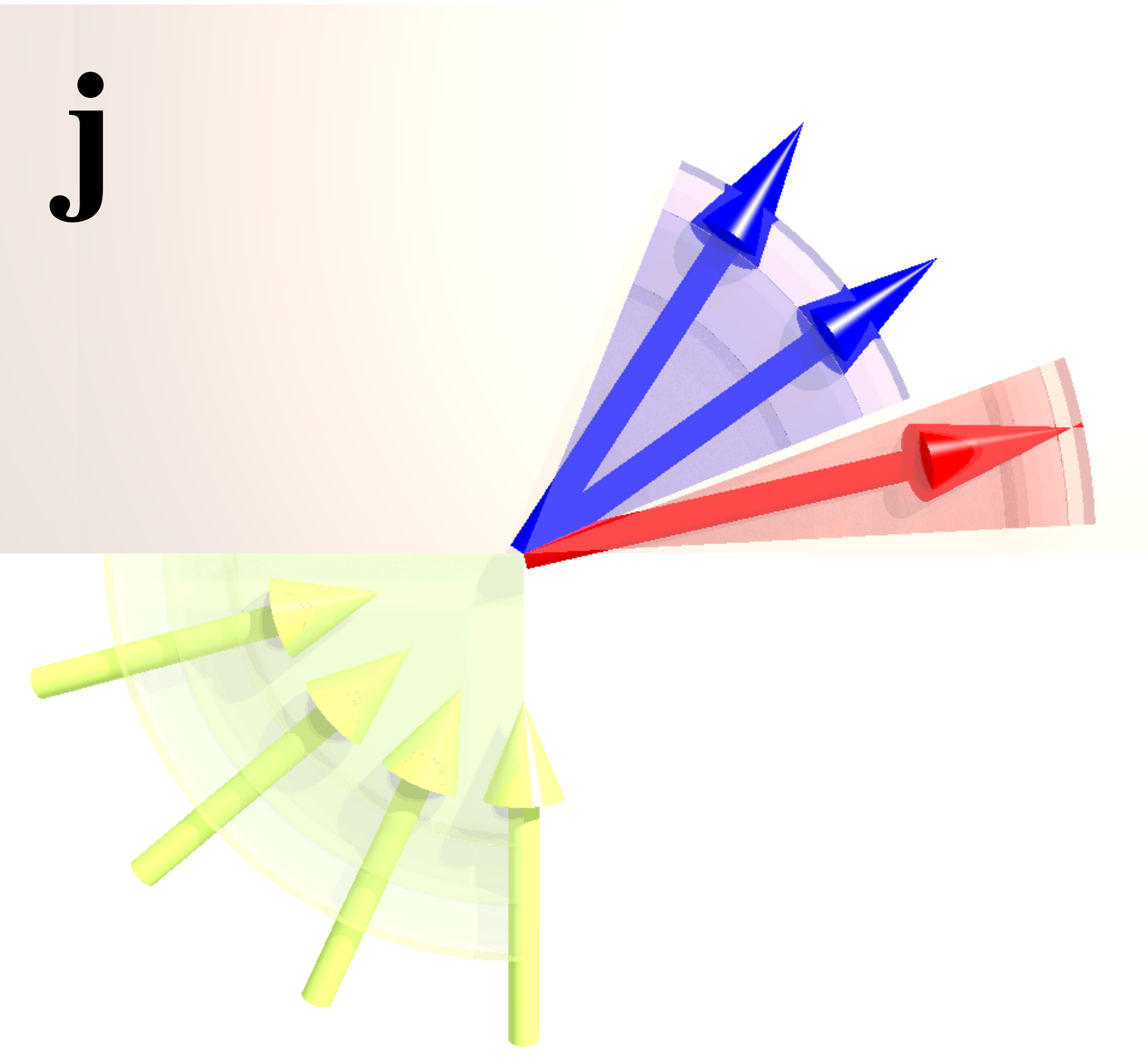}
\end{figure}

\newpage
\noindent
{\small 
Figure 2. 
Optical properties in the hyperbolic regime for 
two geometries, the interface perpendicular to the Rashba field 
$\alphav$ (a,c,e,g,i) and the Rashba field within the interface plane (b,d,f,h,j). 
a,b: Schematic illustration of the geometries. 
The incident wave vector, the wave vector in the Rashba medium and the Poynting vector are denoted by $\qv_{\rm in}$, $\qv$ and  $\Sv$, respectively.
\label{FIGRashba_light}
c,d: 
Dispersion curves for a fixed angular frequency and the relation between the  wave vector $\qv$ and the group velocity $\vv_{\rm g}$. It is seen that the group velocity is refracted to the negative direction in the case c, while the wave vector $\qv$ has a negative $\Re~q_x$ component (backward wave) in the case d.  
\label{FIGdispersion}
e,f: 
Transmittance $T$ (solid lines) and angle of refraction for the Poynting vector $\theta_{S}$ (dotted lines) plotted  as functions of the  angle of incidence $\theta_{\rm in}$ 
for several values of $\tilde\omega(\equiv\omega/\omegaP)=0.78,0.88,0.98$, corresponding to close to the two edges and the center of hyperbolic regime.
g,h: 
Transmittance $T$ plotted as function of $\theta_{S}$.
Focusing effect and its dependence on the frequency are schematically shown in i,j.
Focusing effect is stronger for larger wavelength (shown by red in i and j).
\label{FIGTzvsThetaout}
}

\newpage
\noindent
{\bf Figure 3}

\begin{figure}[H]\centering
\includegraphics[width=0.48\hsize]{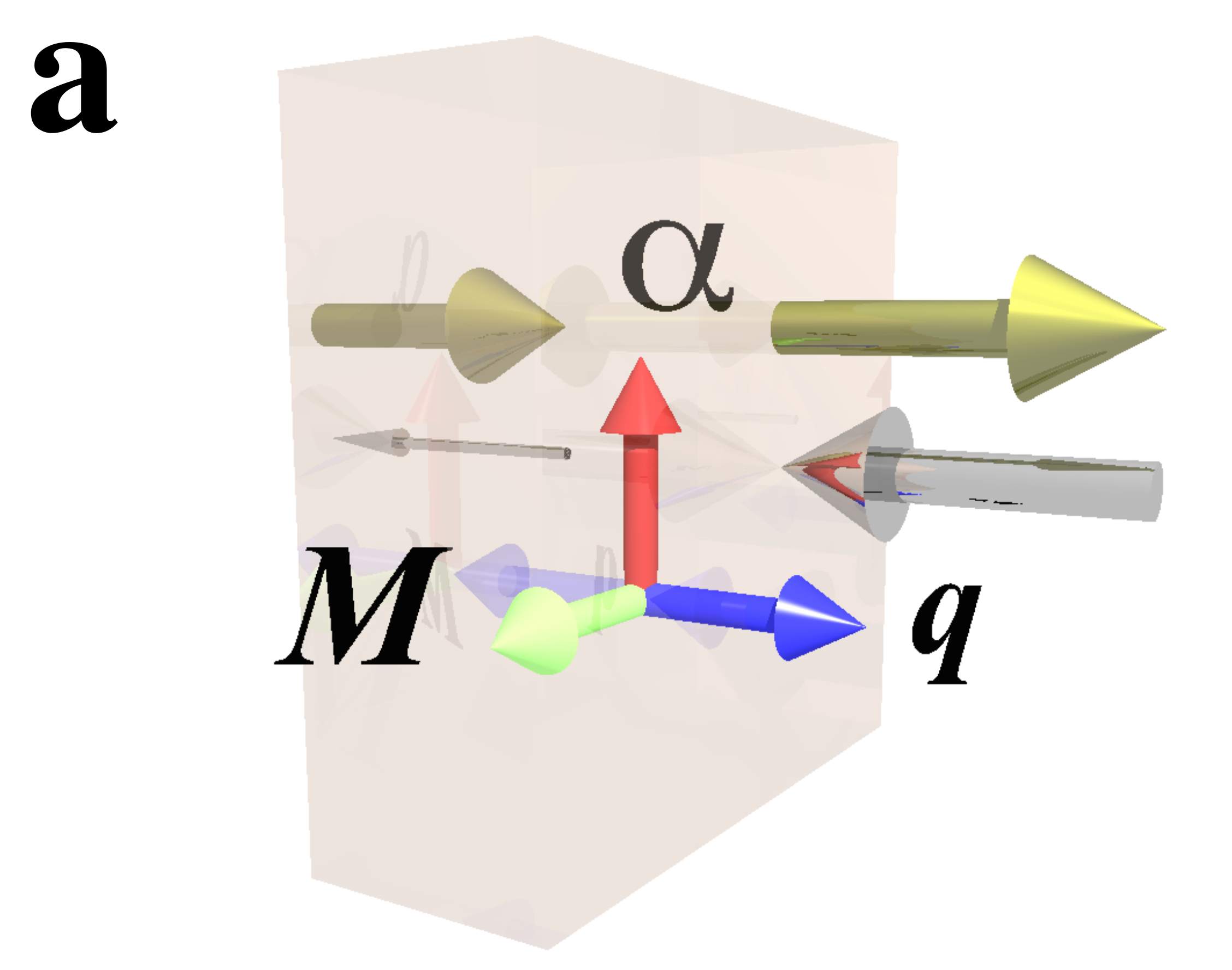}
\includegraphics[width=0.48\hsize]{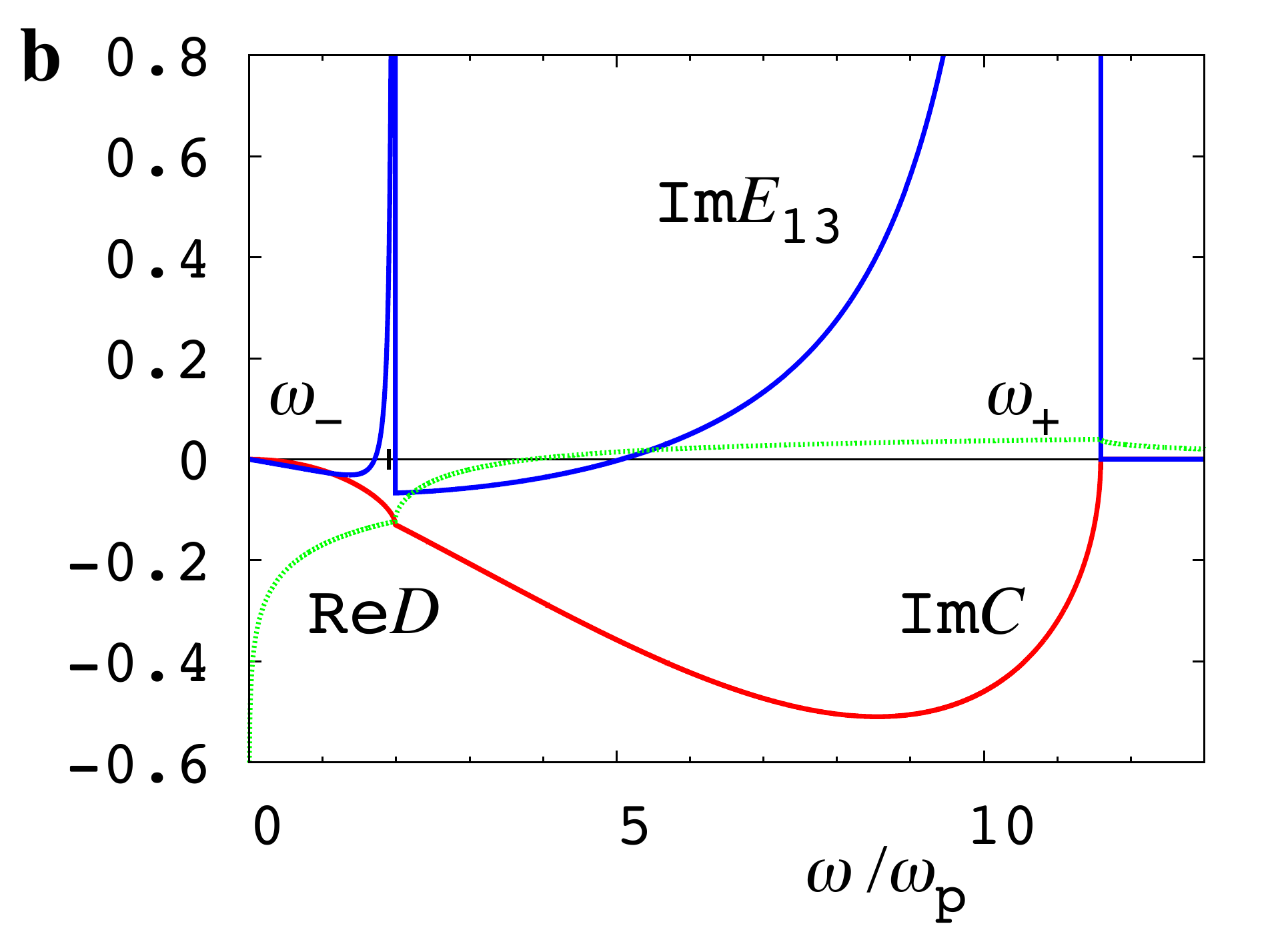}
\end{figure}
\noindent
{\small 
Figure 3. 
a: Schematic illustration showing the configuration of the Rashba field $\alphav$, magnetization $\Mv$ and wave vector $\qv$ required for the directional dichroism 
to appear.
The transmittance depends on the orientation of $\qv$ with respect to 
$\AR=(\alphavhat\times \hat\Mv)$.
b: The imaginary parts of $C(\omega)$ and $E_{13}(\omega)$ (representing the dichroism) and the real part of $D(\omega)$ (representing the strength of the Kerr effect). 
The strength of the dichroism has a singularity at the transition edges $\omega_\pm$, whereas  the Kerr effect has only a cusp.
\label{FIGdichroism}}

\newpage

\noindent
{\bf  \large Acknowledgements}\\
The author would like to thank S. Seki, N. Ogawa, R. Arita, H. Kawaguchi and N. Nagaosa for their valuable comments.
This work was supported by a Grant-in-Aid for Scientific Research (C) (Grant No. 25400344), (A) (Grant No. 24244053) from the Japan Society for the Promotion of Science and  
Grants-in-Aid for Scientific Research on Innovative Areas (Grant No.26103006) from The Ministry of Education, Culture, Sports, Science and Technology (MEXT), Japan. 
The work of J. S. was supported by the INOUE ENRYO Memorial Foundation for Promoting Sciences in Toyo University. \\

\noindent
{\bf \large Author contributions}\\
All authors performed calculations and commented on the manuscript. 
Most of calculations was performed by J.S and A.T. with help from other authors.
J. S. and G. T. contributed to interpretation. 
G. T. planned the investigation and prepared the manuscript. \\

\noindent
{\bf \large Author information}\\
Correspendence should be addressed to J. S. (e-mail: j\_shibata@toyo.jp) and 
G. T. (e-mail: gen.tatara@riken.jp). \\

\noindent
{\bf \large Competing financial interests}\\
The authors declare no competing financial interests.

\newpage

\renewcommand{\theequation}{S.\arabic{equation}}
\setcounter{section}{0}
\setcounter{equation}{0}
\makeatletter
\renewcommand{\@biblabel}[1]{[#1]}
\renewcommand{\@cite}[2]{[{#1\if@tempswa #2\fi}]}
\makeatother

\noindent
{\bf \large Supplementary Information}\\

In this Supplementary Information, we present details on the calculations, 
which is relevant to the discussion of the theory in the main text.

\section{Model Hamiltonian of the Magnetic Rashba system}
In this section, we present the microscopic model Hamiltonian. 
Let us consider a bulk Rashba-electron conductor 
coupled with the magnetization, where 
the Rashba spin$-$orbit interaction breaks the space inversion symmetry and 
the magnetic coupling breaks the time reversal symmetry. 
The Hamiltonian of the conduction electron is given by 
\begin{align}
H &= \sum_{\kv}c^{\dagger}_{\kv}
\left[
\epsilon_{\kv}-\epsilon_{\rm F}+
{\bm \alpha}\cdot(
{\bm \sigma}\times{\kv})
-J_{sd}{\bm M}\cdot{\bm \sigma}\right]c_{\kv}, \nonumber\\
& = \sum_{\kv}c_{\kv}^{\dagger}\left[
\epsilon_{\kv}-\epsilon_{\rm F}
+{\bm \gamma}_{\kv,\Mv}\cdot{\bm \sigma}
\right]c_{\kv}, 
\label{H0}
\end{align}
where $c^{\dagger}_{\kv} = 
(c^{\dagger}_{\kv,\uparrow},c^{\dagger}_{\kv,\downarrow})$ 
is the electron creation operator with wave vector $\kv$, 
the up-arrow(down-arrow)$\uparrow(\downarrow)$ 
represents the spin component along the $z$-axis in the spin space, 
$\epsilon_{\kv} = {\hbar^2\kv^2}/({2m})$ is the energy dispersion 
of free electrons, 
$\epsilon_{\rm F}$ is the Fermi energy, 
${\bm \sigma}$ is a vector of Pauli spin matrices, 
${\bm M}$ is the magnetization vector and 
$J_{sd}$ represents the strength of exchange interaction 
and 
${\bm \gamma}_{\kv,\Mv} = {\bm \gamma}_{\kv} - {\bm M}$ 
with ${\bm \gamma}_{\kv} = \kv \times {\bm \alpha}$. 
The energy dispersion of the Hamiltonian in equation (\ref{H0}) is given by 
\begin{align}
\epsilon_{\kv,\Mv}^{\sigma} = \epsilon_{\kv} + \sigma \gamma_{\kv,\Mv}, 
\end{align} 
where $\sigma = \pm 1$ correspond to the 
spin splitting upper ($\sigma = 1$) and lower ($\sigma = -1$) bands, 
respectively,  
and 
\begin{align}
\gamma_{\kv,\Mv} 
&=\sqrt{{\bm \alpha}^{2}\kv_{\perp}^2-2J_{sd}
({\bm \alpha}\times\Mv)\cdot\kv_{\perp} + J_{sd}^2\Mv\cdot\Mv}. 
\label{gkM}
\end{align} 
where $\kv^{\perp}=\kv-(\kv\cdot\hat{\bm \alpha})\hat{\bm \alpha}$ 
with $\hat{\bm \alpha}={\bm \alpha}/|{\bm \alpha}|$ 
being the unit vector in the direction of Rashba field. 
When ${\bm \alpha}\times \Mv=0$ 
the energy dispersion is symmetric with respect to 
$\kv_{\perp}$-direction in $\kv$-space. 
On the other hand, when ${\bm \alpha}\perp \Mv\neq 0$, the energy dispersion becomes asymmetric in the $\kv_{\perp}$-direction. 
This asymmetry due to the magnetic coupling implies that the light  absorption for electronic interband transitions depends on whether the $\kv_{\perp}$ is parallel or antiparallel to 
${\bm \alpha}\times \Mv$, which will lead to the directional dichroism for electromagnetic wave propagation in the Rashba system.

The one-particle Green's function (resolvent) is given by 
\begin{align}
\label{G1}
G_{\kv,\Mv}(z) &= (z-\epsilon_{\kv}+\epsilon_{\rm F}-{\bm \gamma}_{\kv,\Mv}\cdot{\bm \sigma})^{-1}\nonumber\\
&=\frac{1}{2}\sum_{\sigma=\pm 1}
\frac{1+\sigma\hat{\bm \gamma}_{\kv,\Mv}\cdot{\bm \sigma}}
{z-\epsilon^{\sigma}_{\kv,\Mv} + \epsilon_{\rm F}}, 
\end{align}
where 
$\hat{\bm \gamma}_{\kv,\Mv} = {\bm \gamma}_{\kv,\Mv}/\gamma_{\kv,\Mv}$. 
The retarded and advanced Green's function are given by respectively, 
$G^{\rm R}_{\kv,\Mv}(\epsilon) = G_{\kv,\Mv}(\epsilon+i0^{+})$ and 
$G^{\rm A}_{\kv,\Mv}(\epsilon) = G_{\kv,\Mv}(\epsilon-i0^{+})$, where 
$0^{+}$ is a positive infinitesimal. 
In this paper, 
we consider a clean limit and use the following relation in the latter calculations: 
\begin{align}
\label{R-A}
G^{\rm A}_{\kv,\Mv}(\epsilon) - G^{\rm R}_{\kv,\Mv}(\epsilon) =
i\pi \sum_{\sigma = \pm 1}
\delta(\epsilon-\epsilon^{\sigma}_{\kv,\Mv} +\epsilon_{\rm F})(1 + \sigma \hat{\bm \gamma}_{\kv,\Mv}
\cdot{\bm \sigma}), 
\end{align}
where $\delta(x)$ is the delta function.

\section{Charge and spin response}

Let us evaluate current and spin response 
induced by an external electromagnetic field 
based on the linear response theory. 
The perturbative Hamiltonian representing the linear coupling 
between current and spin of electrons and an external electromagnetic field 
is given by 
\begin{align}
H_{\rm int}(t) 
&= -\int d\rv~\left\{
\hat{\bm j}(\rv)\cdot {\bm A}(\rv,t)-\hbar\gamma\hat{\bm s}(\rv)\cdot
{\bm B}(\rv,t)
\right\}, 
\end{align}
where $\hat{\bm j}(\rv)$ is the current density operator, 
$\hat{\bm s}(\rv)$ is the spin density operator, 
${\bm A}(\rv,t)$ is a vector potential 
yielding the electric field ${\bm E}(\rv,t) = -\dfrac{\partial}{\partial t}{\bm A}(\rv,t)$ 
and the magnetic field ${\bm B}(\rv,t ) = \nabla \times {\bm A}(\rv, t)$,  
and $\gamma (= e/(2m))$ is gyromagnetic ratio. 
In the following calculations, 
we evaluate the expectation values of $\hat{\bm j}(\rv)$ 
and $\hat{\bm s}(\rv)$ in the Fourier space, 
whose components are given by 
\begin{align}
\hat{\bm j}_{\qv} 
&= -e\sum_{\kv}c^{\dagger}_{\kv_{-}}
\left(
{\bm v} + {\bm \alpha}\times{\bm \sigma}/\hbar
\right)c_{\kv_{+}}, \\
\hat{\bm s}_{\qv} &= \sum_{\kv}c^{\dagger}_{\kv_{-}}
{\bm \sigma}c_{\kv_{+}},
\end{align}
where ${\bm v} =\hbar\kv/ m$ is the conventional velocity, 
${\bm \alpha}\times{\bm \sigma}/\hbar$ is the anomalous velocity 
due to the RSOI, which is related to the spin density operator 
and $\kv_{\pm} = \kv\pm\qv/2$. 
Fourier components of the electromagnetic field are also given by 
using ${\bm A}(\qv,\omega) \left(= \int_{-\infty}^{\infty}dt \int d\rv 
e^{-i\qv\cdot \rv + i\omega t}{\bm A}(\rv,t)\right)$ as 
${\bm E}(\qv,\omega) = i\omega {\bm A}(\qv,\omega)$
and 
${\bm B}(\qv,\omega) = i\qv \times {\bm A}(\qv,\omega)$, 
respectively.

The expectation values of  
$\hat{\bm j}_{\qv}$ and $\hat{\bm s}_{\qv}$ 
are written on the linear response of 
${\bm E}(\qv,\omega)$ and ${\bm B}(\qv,\omega)$ as  
\cite{AGD} 

\begin{align}
\langle \hat{j}_{\qv,i}\rangle_{\omega} 
&= \frac{-ie^2}{\omega+i0^{+}}
\left[
\chi_{jj}^{ij}(\qv,\omega;\Mv)-\frac{n_{\rm e}}{m}
\delta_{ij}
\right]E_{j}(\qv,\omega) + e\hbar\gamma
\chi_{js}^{ij}(\qv,\omega;\Mv) B_{j}(\qv,\omega), 
\label{j}\\
\langle \hat{s}_{\qv,i} \rangle_{\omega} 
&= \frac{ie}{\omega+i0^{+}}\chi_{sj}^{ij}(\qv,\omega;\Mv)
E_{j}(\qv,\omega)
-\hbar\gamma \chi_{ss}^{ij}(\qv,\omega;\Mv)B_{j}(\qv,\omega), 
\label{s}
\end{align}
where $e>0$ is the elementary charge and 
$n_{\rm e}$ is an equilibrium electron density and
${\chi}_{jj}^{ij}$, ${\chi}_{js}^{ij}$,  ${\chi}_{sj}^{ij}$
and ${\chi}_{ss}^{ij}$ are current$-$current, current$-$spin, spin$-$current and spin$-$spin 
correlation functions, respectively. 
Here and hereafter, 
repeated indices generally imply summation $(i,j =1,2,3)$.   
These are written by using the Green's function as 
\begin{align}
&\chi_{jj}^{ij}(\qv,\omega;\Mv)=
-\sum_{\kv}\int_{-\infty}^{\infty}
\frac{d\vare}{2\pi i}
{\rm tr}\left[
\tilde{v}_{i}G_{\kv_{+},\Mv}(\vare_{+})\tilde{v}_{j}G_{\kv_{-},\Mv}(\vare_{-})
\right]^{<}, 
\label{cjj}
\\
&\chi_{js}^{ij}(\qv,\omega;\Mv)=
-\sum_{\kv}\int_{-\infty}^{\infty}
\frac{d\epsilon}{2\pi i}
{\rm tr}\left[
\tilde{v}_{i}G_{\kv_{+},\Mv}(\epsilon_{+}){\sigma}_{j}G_{\kv_{-},\Mv}(\epsilon_{-})
\right]^{<}, 
\label{cjs}
\\
&\chi_{sj}^{ij}(\qv,\omega;\Mv) = -\sum_{\kv}\int_{-\infty}^{\infty}
\frac{d\epsilon}{2\pi i}
{\rm tr}\left[
{\sigma}_{i}G_{\kv_{+},\Mv}(\epsilon_{+})
\tilde{v}_{j}G_{\kv_{-},\Mv}(\epsilon_{-})
\right]^{<}, 
\label{csj}
\\
&\chi_{ss}^{ij}(\qv,\omega;\Mv) = 
-\sum_{\kv}\int_{-\infty}^{\infty}
\frac{d\epsilon}{2\pi i}
{\rm tr}\left[
{\sigma}_{i}G_{\kv_{+},\Mv}(\epsilon_{+})
{\sigma}_{j}G_{\kv_{-},\Mv}(\epsilon_{-})
\right]^{<}. 
\label{css}
\end{align}
where $\tilde{\bm v} = {\bm v} + {\bm \alpha}\times{\bm \sigma}/\hbar$, 
$\kv_{\pm} = \kv \pm \qv/2$ and 
$\epsilon_{\pm} = \epsilon \pm \omega/2$. 
The Green's function $G_{\kv,\Mv}(\epsilon)$ here is a path-ordered one \cite{Rammer86}. 
The lesser component of $G_{\kv,\Mv}(\epsilon)$ is given by \cite{{Langreth76},{Haug98}} 
\begin{align}
\label{lesser}
G^{<}_{\kv,\Mv}(\epsilon) = f(\epsilon)\left(
G^{\rm A}_{\kv,\Mv}(\epsilon)-G^{\rm R}_{\kv,\Mv}(\epsilon)
\right), 
\end{align}
where $f(\epsilon)=(1+e^{\epsilon/T})^{-1}$ is the Fermi distribution function 
with $T$ being the temperature.

The key to understanding the optical magnetoelectric properties is to evaluate the 
optical conductivity, $\sigma_{ij}(\qv,\omega;\Mv)$, which can be related to
the correlation functions as 
 \begin{align}
 \sigma_{ij}(\qv,\omega) 
 = \frac{-ie^2}{\omega+i0^{+}}\left[\chi_{jj}^{ij}-\frac{n_{\rm e}}{m}\delta_{ij}\right]
 +\frac{e\hbar\gamma}{\omega+i0^{+}}\left[
 \chi_{js}^{il}\vare_{lkj} 
 + \vare_{ikl}\chi_{sj}^{lj}
 \right]q_{k}
 +
 \frac{i\hbar^2\gamma^2}{\omega+i0^{+}}\chi_{ss}^{mn}
 \vare_{ikm}\vare_{jln}q_{k}q_{l}, 
  \label{oc1}
 \end{align}
 where 
 $\vare_{ijk}$ is the totally anti-symmetric tensor with $\vare_{123}=1$. 
In the main text, the optical conductivity $\sigma_{ij}(\qv,\omega;\Mv)$ 
is evaluated to the  first order in $\qv$ and $\Mv$. 
Thus $\chi_{jj}^{ij}(\qv,\omega;\Mv)$ 
may be expanded up to the linear in $\qv$ and $\Mv$. 
For $\chi_{js}^{ij}(\qv,\omega;\Mv)$ 
and $\chi_{sj}^{ij}(\qv,\omega;\Mv)$, 
we can put $\qv=0$ and evaluate these up to the first order in $\Mv$. 
Since the last term in equation (\ref{oc1}) is the second order in $\qv$, 
this is negligible {\color{black} in the optical conductivity.}

\section{Calculation of optical conductivity tensor}

In this section, 
we present calculation results of correlation functions, $\chi_{jj}^{ij}$, $\chi_{js}^{ij}$, $\chi_{sj}^{ij}$ and $\chi^{ss}_{ij}$ and the optical conductivity tensor, $\sigma_{ij}$,  
and give the explicit form of coefficients, 
$C$, $D$, $E_{i}~(i=1,2,3)$, 
of $\sigma^{M}_{ij}$ in the main text.   

Let us first perform the $\epsilon$-integral in equations (\ref{cjj})-(\ref{css}).  
Following the Langreth's method \cite{{Langreth76},{Haug98}}, 
one can calculate 
the lesser component of $G(\epsilon_{+})G(\epsilon_{-})$ as 
\begin{align}
\left[G(\epsilon_{+})G(\epsilon_{-})\right]^{<} &= 
G^{<}(\epsilon_{+})G^{\rm A}(\epsilon_{-}) + G^{\rm R}(\epsilon_{+})G^{<}(\epsilon_{-})\nonumber\\
&= 
f(\epsilon_{+})(G^{\rm A}(\epsilon_{+})-G^{\rm R}(\epsilon_{+}))G^{\rm A}(\epsilon_{-}) + f(\epsilon_{-})G^{\rm R}(\epsilon_{+})(G^{\rm A}(\epsilon_{-})-G^{\rm R}(\epsilon_{-})). 
\end{align}
Applying this rule to the correlation functions of equations (\ref{cjj})-(\ref{css}) 
and using equation (\ref{R-A}), we can carry out the $\epsilon$-integral due to the 
delta function. 
The next step is to expand correlation function with respect to $\qv$ and $\Mv$ 
and subsequently to carry out the integration with respect to the rotation angle 
around the ${\bm \alpha}$-axis. Since these calculations are rather complicated, 
we here only show the results as  
\begin{align}
\chi_{jj}^{ij}&\simeq
-\frac{n_{\rm e}}{m}C(\omega)\delta^{\perp}_{ij}
 -i\frac{n_{\rm e}}{m}
 \frac{J_{sd}\hbar\omega}{\epsilon_{\rm F}^2}
 D(\omega)
 (\hat{\bm \alpha}\cdot\Mv)
\vare_{ijk}\hat{\alpha}_{k}\nonumber\\
&~~~~+\frac{n_{\rm e}}{m}\frac{J_{sd}}{k_{\rm F}\epsilon_{\rm F}}
\left[
E_{1}(\omega)(\hat{\bm \alpha}\times\Mv ) \cdot \qv \delta^{\perp}_{ij}
+\frac{1}{2}E_{2}(\omega) \left\{
(\hat{\bm \alpha}\times \Mv)_{i}q^{\perp}_{j} + 
q^{\perp}_{i}(\hat{\bm \alpha}\times\Mv)_{j}
\right\}
\right]
\label{chi-jj-f}\\
\chi_{js}^{ij} 
&\simeq
\frac{\hbar}{\alpha}\frac{n_{\rm e}}{m}C(\omega)\vare_{ijk}\hat{\alpha}_{k}
+i\frac{\hbar}{\alpha}\frac{n_{\rm e}}{m}\frac{J_{sd}\hbar\omega}{\epsilon_{\rm F}^2}
\left(N_{i}(\omega)\hat{\alpha}_{j}
-\hat{\bm \alpha}\cdot{\bm N}(\omega)\delta_{ij}\right),
\label{chi-js-f}\\
\chi_{sj}^{ij} 
&\simeq-\frac{\hbar}{\alpha}\frac{n_{\rm e}}{m}C(\omega)\vare_{ijk}\hat{\alpha}_{k}
-i\frac{\hbar}{\alpha}\frac{n_{\rm e}}{m}\frac{J_{sd}\hbar\omega}{\epsilon_{\rm F}^2}
\left(\hat{\alpha}_{i}N_{j}(\omega)
-\hat{\bm \alpha}\cdot{\bm N}(\omega)\delta_{ij}\right),
\label{chi-sj-f}\\
\chi_{ss}^{ij}&\simeq \frac{\hbar}{\alpha}\frac{n_{\rm e}}{m}C(\omega)
(\delta^{\perp}_{ij}-2\delta_{ij})-i\frac{\hbar^2}{\alpha^2}
\frac{n_{\rm e}}{m}\frac{J_{sd}\hbar\omega}{\epsilon_{\rm F}^2}
\vare_{ijk}N_{k}(\omega)
\end{align}
where $\delta^{\perp}_{ij} = \delta_{ij}-\hat{\alpha}_{i}\hat{\alpha}_{j}$ 
with $\hat{\bm \alpha}={\bm \alpha}/|{\bm \alpha}|$, 
${\bm N}(\omega) = 
D(\omega)(\Mv\cdot\hat{\bm \alpha})\hat{\bm \alpha}
+\tilde{\alpha}\dfrac{\epsilon_{\rm F}}{\hbar\omega}E_{3}(\omega)
\left(\Mv-(\Mv\cdot\hat{\bm \alpha})\hat{\bm \alpha}\right)$ 
and the frequency-dependent coefficients 
are defined as 
\begin{align}
&C(\omega) = 
-\frac{4\tilde{\alpha}^2}{n_{\rm e}}\epsilon_{\rm F}
\sum_{\kv}\frac{\gamma_{\kv}s_{\kv}}{H_{\kv}}
\label{B-C}
\\
&D(\omega) =  
-\frac{4\tilde{\alpha}^2}{n_{\rm e}}\epsilon_{\rm F}^3
\sum_{\kv}\frac{s_{\kv}}{\gamma_{\kv}H_{\kv}}
\label{B-D}\\
&E_{1}(\omega)=
\frac{1}{4n_{\rm e}\tilde{\alpha}}\frac{\epsilon_{\rm F}}{\hbar\omega}
\sum_{\kv}\left(
\gamma_{\kv}s'_{\kv}+\frac{m\alpha^2}{\hbar^2}n'_{\kv}
\right)\nonumber\\
&\qquad~~
+\frac{\tilde{\alpha}}{2n_{\rm e}}\epsilon_{\rm F}^2\hbar\omega 
\sum_{\kv}
\left\{\frac{s_{\kv}}{\gamma_{\kv}H_{\kv}}
-12\left(
 \frac{2\gamma^2_{\kv}n'_{\kv}}{H_{\kv}^2}
+\frac{\gamma_{\kv}s_{\kv}}{H_{\kv}^2}
+16\frac{\gamma_{\kv}^3 s_{\kv}}{H_{\kv}^3}
\right)
\right\}
\label{B-E1}\\
&E_{2}(\omega)=
\frac{1}{4n_{\rm e}\tilde{\alpha}}\frac{\epsilon_{\rm F}}{\hbar\omega}
\sum_{\kv}\left(
\gamma_{\kv}s'_{\kv}+\frac{m\alpha^2}{\hbar^2}n'_{\kv}
\right)\nonumber\\
&\qquad~~
+\frac{\tilde{\alpha}}{2n_{\rm e}}\epsilon_{\rm F}^2\hbar\omega 
\sum_{\kv}
\left\{
\frac{s_{\kv}}{\gamma_{\kv}H_{\kv}}
+4\left( \frac{2\gamma^2_{\kv}n'_{\kv}}{H_{\kv}^2}
+\frac{\gamma_{\kv}s_{\kv}}{H_{\kv}^2}
+16\frac{\gamma_{\kv}^3 s_{\kv}}{H_{\kv}^3}\right)
\right\}
\label{B-E2}\\
&E_{3}(\omega) = 
-\frac{2\tilde{\alpha}}{n_{\rm e}}\epsilon^{2}_{\rm F}\hbar\omega
\sum_{\kv}\left(
\frac{s_{\kv}}{\gamma_{\kv}H_{\kv}}
+\frac{n'_{\kv}}{H_{\kv}}+\frac{8\gamma_{\kv}s_{\kv}}{H_{\kv}^{2}}
\right).
\label{B-E3} 
\end{align}
with 
$\tilde{\alpha} = \dfrac{m\alpha}{\hbar^2k_{\rm F}}$ being the dimensionless Rashba 
strength parameter, 
$H_{\kv} = (\hbar\omega+i0^{+})^2-4\gamma_{\kv}^2$ and 
$s_{\kv} = \sum_{\sigma=\pm}\sigma f_{\kv\sigma}$, 
$n_{\kv} = \sum_{\sigma=\pm} f_{\kv\sigma}$, 
$s'_{\kv} = \sum_{\sigma=\pm}\sigma f'_{\kv\sigma}$, 
$n'_{\kv} = \sum_{\sigma=\pm}f'_{\kv\sigma}$ 
with $f_{\kv\sigma} = f(\epsilon_{\kv}+\sigma\gamma_{\kv}-\epsilon_{\rm F})$ 
and $f'_{\kv\sigma} = \left.\dfrac{df(\epsilon)}{d\epsilon}\right|_{\epsilon=\epsilon_{\kv}+\sigma\gamma_{\kv}-\epsilon_{\rm F}}$.

Substituting equations (\ref{chi-jj-f}), (\ref{chi-js-f}) and (\ref{chi-sj-f}) into the equation (\ref{oc1}), 
we obtain the optical conductivity as 
\begin{align}
 \sigma_{ij}(\qv,\omega;\Mv)
&=
\dfrac{ie^2}{\omega+i0^{+}}\frac{n_{\rm e}}{m}\left(
\delta_{ij}(1+C(\omega))-\hat{\alpha}_{i}\hat{\alpha}_{j}C(\omega)
\right)
+i\hbar\gamma \kappa_{\rm E}(\omega)(q_{i}\hat{\alpha}_{j}-\hat{\alpha}_{i}q_{j})
+\sigma^{M}_{ij}(\qv,\omega)
\label{op2-1}
\end{align}
where 
$\kappa_{\rm E} (\omega)= i\dfrac{e\hbar n_{\rm e}}{m\alpha}\dfrac{C(\omega)}{\omega}$ is the Rashba$-$Edelstein coefficient and 
\begin{align}
\sigma_{ij}^{M} (\qv,\omega)
=&-\frac{e^2n_{\rm e}}{m}\frac{\hbar}{\epsilon_{\rm F}^2}
D(\omega)(\hat{\bm \alpha}\cdot\Mv)
\left[
\vare_{ijk}\hat{\alpha}_{k}
-\frac{i}{2\tilde{\alpha}k_{\rm F}}(
\hat{\alpha}_{i}\vare_{jkl}
+\vare_{ikl}\hat{\alpha}_{j}
)\hat{\alpha}_{k}q_{l}
\right]\nonumber\\
&-i\frac{e^2n_{\rm e}}{m}\frac{1}{k_{\rm F}\epsilon_{\rm F}\omega}
\left[
E_{1}(\omega)(\hat{\bm \alpha}\times\Mv ) \cdot \qv \delta^{\perp}_{ij}
+
\frac{1}{2}E_{2}(\omega) \left\{
(\hat{\bm \alpha}\times \Mv)_{i}q^{\perp}_{j} + 
q^{\perp}_{i}(\hat{\bm \alpha}\times\Mv)_{j}
\right\}\right.\nonumber\\
&\left.-
\frac{1}{2}E_{3}(\omega)
\left[
M^{\perp}_{i}(\hat{\bm \alpha}\times \qv)_{j} + (\hat{\bm \alpha}\times\qv)_{i}M^{\perp}_{j}
\right]\right]
\label{sigma-M}
\end{align}
with $\qv^{\perp}=\qv-(\qv\cdot\hat{\bm \alpha})\hat{\bm \alpha}$ 
and $\Mv^{\perp}=\Mv-(\Mv\cdot\hat{\bm \alpha})\hat{\bm \alpha}$. 
The first term of equation (\ref{op2-1}) leads to 
the softening of plasma frequency in the directions perpendicular to 
$\hat{\bm \alpha}$. 
The second term represents the direct and inverse Edelstein effects, 
which leads to the natural optical activity (see main text). 
The effect of the magnetization is contained in $\sigma^{M}_{ij}$. 
The term proportional to $(\hat{\bm \alpha}\cdot \Mv)$ of equation (\ref{sigma-M}) 
gives an anomalous Hall effect, which leads to a magneto-optical response (Kerr effect). 
Other terms including $\hat{\bm \alpha}\times\Mv$ or ${\bm M}^{\perp}$ 
and having diagonal components linear in the wave vector $\qv$
of equation (\ref{sigma-M}) induce directional dichroism.

\section{Wave propagation in the Rashba sytem}
In this section, we consider the wave propagation in the Rashba system 
based on the optical conductivity of equation (\ref{op2-1}) 
and lead to the dispersion relation of equation (11) in the main text. 

When we choose 
$\hat{\bm \alpha} = (0,0,1)$, $\Mv = (0,M,0)$ and $\qv = (q_{x},0,q_{z})$, 
the optical conductivity tensor can be written as 
\begin{align}
\sigma_{ij} = \left(
    \begin{array}{ccc}
      (1+C)\sigma_{\rm D} +i\frac{e^2n_{\rm e}}{m}\frac{J_{sd}M}{k_{\rm F}\epsilon_{\rm F}\omega}E_{12}q_{x} 
      & 0 
      & i\hbar\gamma \kappa_{\rm E}q_{x}\\
      0 
      & (1+C)\sigma_{\rm D} +i\frac{e^2n_{\rm e}}{m}\frac{J_{sd}M}{k_{\rm F}\epsilon_{\rm F}\omega}E_{13}q_{x} 
      & 0\\
      -i\hbar\gamma \kappa_{\rm E}q_{x}
      & 0 
      & \sigma_{\rm D}
    \end{array}
  \right), 
\end{align}
where $E_{12} = E_{1}+E_{2}$ and $E_{13}=E_{1}+E_{3}$. 
Thus the dielectric tensor $\vare_{ij} = \delta_{ij} 
+ \dfrac{i}{\vare_{0}\omega}\sigma_{ij}$ is given by 
\begin{align}
\vare_{ij} = \left(
    \begin{array}{ccc}
     \tilde{\vare}_{x}
      & 0 
      & -\tilde{\beta} \\
      0 
      & \tilde{\vare}_{y} 
      & 0\\
      \tilde{\beta}
      & 0 
      & \vare_{z}
    \end{array}
  \right)
  \label{dt1}
\end{align}
where 
\begin{align}
&\tilde{\vare}_{x} = \vare_{x}
-\frac{\omega_{\rm p}^2}{\omega^2}\frac{J_{sd}M}{k_{\rm F}\epsilon_{\rm F}}(E_{1}+E_{2}) q_{x},\\
&\tilde{\vare}_{y} = \vare_{x}
-\frac{\omega_{\rm p}^2}{\omega^2}\frac{J_{sd}M}{k_{\rm F}\epsilon_{\rm F}}\left(E_{1}+E_{3}\right) q_{x},\\
&\vare_{z} = 1-\frac{\omega^{2}_{\rm p}}{\omega(\omega+i\eta)},\\
&\vare_{x} = 1-\frac{\omega^{2}_{\rm p}}{\omega(\omega+i\eta)}(1+C),\\
&\tilde{\beta}= \frac{\mu_{0}c^2}{\omega}\beta
\end{align}
with $\beta = \hbar \gamma \kappa_{\rm E} q_{x}$. 
In general, the electromagntic wave propagation in conducting media 
is described by the following equation:\cite{Landau}
\begin{align}
\label{weq-0}
\left[
c^2(q^2\delta_{ij}-q_{i}q_{j})-\omega^2\vare_{ij}
\right]E_{j}=0, 
\end{align}
where 
$c$ is the light velocity. 
Substituting equation (\ref{dt1}) into the equation (\ref{weq-0}), 
we obtain the following characteristic equation yielding a dispersion relation for plane wave 
as 
\begin{align}
 \left|
    \begin{array}{ccc}
      c^2q^2_{z}-\omega^2\tilde{\vare}_{x} & 0 & -c^2q_{x}q_{z}-\mu_{0}\beta \omega \\
      0 & c^2(q^2_{x}+q^2_{z})-\omega^2\tilde{\vare}_{y} & 0 \\
     -c^2q_{x}q_{z} +\mu_{0}\beta\omega & 0 & c^2q_{x}^2-\vare_{z}
    \end{array}
  \right|=0. 
\end{align}
Thus for $E_{y}$, 
which represents an ordinary wave, we have the dispersion relation 
\begin{align}
c^2(q_{x}^2+q_{z}^2)-\omega^2 \tilde{\vare}_{y}=0. 
\end{align}
Putting the wave vector on $\qv = q(\sin \theta_{q},0,\cos\theta_{q})$, 
we obtain 
\begin{align}
q \simeq 
\frac{\omega}{c}\sqrt{\vare_{x}}
+\frac{1}{2}\frac{\omega_{\rm p}^2}{c^2}\frac{J_{\rm cd}M}{k_{\rm F}\epsilon_{\rm F}}E_{13}\hat{\qv}\cdot{\cal A}
\label{disp-DD}
\end{align}
when $\dfrac{\omega_{\rm p}}{ck_{\rm F}}\dfrac{J_{sd}M}{\epsilon_{\rm F}}\ll 1$, 
where $\hat{\qv} = \qv/q$ and ${\cal A} = \hat{\bm \alpha}\times \hat{\bm M}$
with $\hat{\Mv} =\Mv/|\Mv|$, 
which leads to equation (11) in the main text. 
Thus we see that the second term including the factor $\hat{\qv}\cdot{\cal A}$ of equation (\ref{disp-DD}), 
which depends on the wave propagating direction $\hat{\qv}$, 
leads to the directional dichroism. 
The difference in the absorption rate of 
positive $(\theta_{q})$ and negative $(\pi-\theta_{q})$  wave propagation is proportional 
to the imaginary part of $E_{13}$ (see main text). 
 
\section{$\kv$-integrals of coefficients, $C$, $D$, $E_{1}$, $E_{2}$ and $E_{3}$}
In this section, we show the results of $\kv$-integrals of equations (\ref{B-C})-(\ref{B-E3}). 

Using 
\begin{align}
&S \equiv \sum_{\kv}\frac{s_{\kv}}{\gamma_{\kv} H_{\kv}}
\label{C-S}\\
&N \equiv \sum_{\kv}\frac{n'_{\kv}}{H_{\kv}}
\label{C-N}
\end{align}
we can write equations (\ref{B-C})-(\ref{B-E3}) as 
\begin{align}
&C(\omega) = C(0) - \frac{\tilde{\alpha}^2}{n_{\rm e}}\epsilon_{\rm F}(\hbar\omega)^2 S
\label{C-C}\\
&D(\omega) = - \frac{4\tilde{\alpha}^2}{n_{\rm e}}\epsilon^3_{\rm F} S
\label{C-D}\\
&E_{1}(\omega) = \frac{\tilde{\alpha}^{-1}}{4n_{\rm e}}\frac{\epsilon_{\rm F}}{\hbar\omega}\sum_{\kv}\left(\gamma_{\kv}s'_{\kv} + \frac{m\alpha^2}{\hbar^2}n'_{\kv}\right)
+\frac{2\tilde{\alpha}}{n_{\rm e}}\epsilon_{\rm F}^2\hbar\omega
\left(
-2S+\frac{3}{2}N-\frac{9}{4}\omega\frac{\partial S}{\partial \omega}
+\frac{3}{4}\omega\frac{\partial N}{\partial \omega}
-\frac{3}{8}\omega^2\frac{\partial^2 S}{\partial \omega^2}
\right)
\label{C-E1}
\\
&E_{2}(\omega) = \frac{\tilde{\alpha}^{-1}}{4n_{\rm e}}\frac{\epsilon_{\rm F}}{\hbar\omega}\sum_{\kv}\left(\gamma_{\kv}s'_{\kv} + \frac{m\alpha^2}{\hbar^2}n'_{\kv}\right)
+\frac{2\tilde{\alpha}}{n_{\rm e}}\epsilon_{\rm F}^2\hbar\omega
\left(
S-\frac{1}{2}N+\frac{3}{4}\omega\frac{\partial S}{\partial \omega}
-\frac{1}{4}\omega\frac{\partial N}{\partial \omega}
+\frac{1}{8}\omega^2\frac{\partial^2 S}{\partial \omega^2}
\right)
\label{C-E2}
\\
&E_{3}(\omega) =
+\frac{2\tilde{\alpha}}{n_{\rm e}}\epsilon_{\rm F}^2\hbar\omega
\left(
S-N+\omega\frac{\partial S}{\partial \omega}
\right)
\label{C-E3}
\end{align}
where 
\begin{align}
C(0) = \frac{\tilde{\alpha}^2}{n_{\rm e}}\epsilon_{\rm F}
\sum_{\kv}\frac{s_{\kv}}{\gamma_{\kv}}.
\end{align}
{\color{black} At zero temperature,} 
we have 
\begin{align}
n_{\rm e} &= -\sum_{\kv}\sum_{\sigma}\delta(\epsilon_{\kv}+\sigma \gamma_{\kv}-\epsilon_{\rm F}) = 
\frac{k_{\rm F}^3}{3\pi^2}(1+2\Delta_{\tilde{\alpha}}), \\
C(0) &= \frac{\tilde{\alpha}^2}{1+2\Delta_{\tilde{\alpha}}}\frac{3\pi^2}{k_{\rm F}^3}\epsilon_{\rm F}
\frac{1}{4\pi^2\alpha}
\int_{0}^{\pi}d\vartheta \int_{0}^{\infty}dk k 
\sum_{\sigma}\sigma \theta(\epsilon_{\kv} + \sigma \gamma_{\kv} -\epsilon_{\rm F})\nonumber\\
&=-\frac{\Delta_{\tilde{\alpha}}}{1+2\Delta_{\tilde{\alpha}}}
\end{align}
where 
\begin{align}
\Delta_{\tilde{\alpha}} = \frac{3}{4}\tilde{\alpha}^2\left(
1+\frac{1+\tilde{\alpha}^2}{\tilde{\alpha}}
\tan^{-1}\tilde{\alpha}\right)
\end{align}

{\color{black} At zero temperature,} 
we write $S$ in equation (\ref{C-S}) and 
$N$ in equation (\ref{C-N}) as 
\begin{align}
&S=\frac{1}{4\pi^2\alpha}\int_{0}^{\pi}d\vartheta \int_{0}^{\infty}dk k 
\sum_{\sigma}\sigma \frac{ \theta(\epsilon_{\kv} + \sigma \gamma_{\kv} -\epsilon_{\rm F})}{(\hbar\omega+i0^{+})^2-4\gamma_{\kv}^2}, \\
&N=-\frac{1}{4\pi^2}\int_{0}^{\pi}d\vartheta\sin\vartheta \int_{0}^{\infty}dk k^2 
\sum_{\sigma} \frac{\delta(\epsilon_{\kv} + \sigma \gamma_{\kv} -\epsilon_{\rm F})}{(\hbar\omega+i0^{+})^2-4\gamma_{\kv}^2}. 
\end{align}
Carrying out the integrals, we obtain 
the real and imaginary parts of $S$ and $N$ as 
\begin{align}
&{\rm Re}S = -\frac{1}{8\pi^2}\frac{1}{\alpha^3}
\left[\frac{1}{\overline{\omega}}\sum_{\lambda=\pm1}
(\lambda-\overline{\omega})L_{\lambda}+2\tan^{-1}\tilde{\alpha}\right], 
\label{RS}\\
&{\rm Im}S
=-\frac{1}{16\pi}\frac{1}{\alpha^3}\frac{1}{\overline{\omega}}
\sum_{\lambda=\pm 1}\lambda S_{\lambda}=
-\frac{1}{16\pi}\frac{1}{\alpha^3}\frac{1}{\overline{\omega}}
\times
\begin{cases}
S_{+}-S_{-}
 & 0<\overline{\omega}<\overline{\omega}_{-}
 \\
 -S_{-}
 &    
\overline{\omega}_{-} <\overline{\omega} < \overline{\omega}_{+}\\
0&       \overline{\omega}_{+}< \overline{\omega}
    \end{cases}
\label{IS}\\
&{\rm Re}N = -
\frac{1}{8\pi^2}\frac{\tilde{\alpha}^2}{\alpha^3}\sum_{\lambda}\frac{\lambda}{\lambda-\overline{\omega}}
R_{\lambda}, 
\label{RN}\\
&    {\rm Im}N = \frac{1}{16\pi}\frac{\tilde{\alpha}^2}{\alpha^3}
\sum_{\lambda=\pm 1}\frac{1}{S_{\lambda}}=
\frac{1}{16\pi}\frac{\tilde{\alpha}^2}{\alpha^3}\times
\begin{cases}
\dfrac{1}{S_{+}}+\dfrac{1}{S_{-}}
 & 0<\overline{\omega}<\overline{\omega}_{-}
 \\
 \dfrac{1}{S_{-}}
 &    
\overline{\omega}_{-} <\overline{\omega} < \overline{\omega}_{+}\\
0&       \overline{\omega}_{+}< \overline{\omega}
    \end{cases}
    \label{IN}
\end{align}
where 
$\overline{\omega}_{\rm F} = \omega/(4\omega_{\rm F}$) with $\hbar\omega=\epsilon_{\rm F}$ 
is a dimensionless angular frequency normalized by 
the Fermi energy, 
\begin{align}
&L_{\lambda}=\sqrt{Q_{\lambda}}\tan^{-1}
\left(\frac{\tilde{\alpha}}{\sqrt{Q_{\lambda}}}\right),\\
&R_{\lambda}=\frac{1}{\sqrt{Q_{\lambda}}}\tan^{-1}
\left(\frac{\tilde{\alpha}}{\sqrt{Q_{\lambda}}}\right),\\
&Q_{\lambda} 
=\frac{\overline{\omega}^2+2\lambda\tilde{\alpha}^2\overline{\omega}-\tilde{\alpha}^2}{(\lambda-\overline{\omega})^2},\\
&S_{\lambda} = \sqrt{\tilde{\alpha}^2-2\lambda\tilde{\alpha}^{2}\overline{\omega}-\overline{\omega}^2},
\end{align}
and 
\begin{align}
\overline{\omega}_{\pm}=\tilde{\alpha}(\sqrt{1+\tilde{\alpha}^2}\pm \tilde{\alpha})
\end{align}
are frequencies at the transition edges. 
The imaginary part of $C(\omega)$ has cusps at $\overline{\omega}_{\pm}$ 
(see equation (\ref{V-IC}) bellow). 
Note that when $Q_{\lambda}<0$, the following replacement is understood 
\begin{align}
&\sqrt{Q_{\lambda}}\tan^{-1}
\left(\frac{\tilde{\alpha}}{\sqrt{Q_{\lambda}}}\right)
~~\to~~
\sqrt{-Q_{\lambda}}\tanh^{-1}\left(\frac{\tilde{\alpha}}{\sqrt{-Q_{\lambda}}}\right) =
\frac{1}{2}\sqrt{-Q_{\lambda}}
{\rm ln}\left|
\frac{\sqrt{-Q_{\lambda}}+\tilde{\alpha}}{\sqrt{-Q_{\lambda}}-\tilde{\alpha}}
\right|
\\
&\frac{1}{\sqrt{Q}_{\lambda}}\tan^{-1}\left(\frac{\tilde{\alpha}}{\sqrt{Q_{\lambda}}}\right)
~~\to~~
-\frac{1}{\sqrt{-Q}_{\lambda}}\tanh^{-1}\left(\frac{\tilde{\alpha}}{\sqrt{-Q_{\lambda}}}\right)
=
-\frac{1}{2}\frac{1}{\sqrt{-Q_{\lambda}}}
{\rm ln}\left|
\frac{\sqrt{-Q_{\lambda}}+\tilde{\alpha}}{\sqrt{-Q_{\lambda}}-\tilde{\alpha}}
\right|
\end{align}

Substituting these results into equations (\ref{B-C}) and (\ref{B-D}), 
we obtain  
\begin{align}
&{\rm Re}C = \frac{3}{4}\frac{\tilde{\alpha}^{-1}}{1+2\Delta_{\tilde{\alpha}}}
\left[\overline{\omega}\sum_{\lambda=\pm1}(\lambda-\overline{\omega})L_{\lambda}+2\overline{\omega}^2
\tan^{-1}\tilde{\alpha}\right] - \frac{\Delta_{\tilde{\alpha}}}{1+2\Delta_{\tilde{\alpha}}},
\label{V-RC}
\\
&{\rm Im}C =\frac{3\pi}{8} \frac{\tilde{\alpha}^{-1}}{1+2\Delta_{\tilde{\alpha}}}
\overline{\omega}\sum_{\lambda}\lambda S_{\lambda},
\label{V-IC}\\
&{\rm Re}D =\frac{3}{16}\frac{\tilde{\alpha}^{-1}}{1+2\Delta_{\tilde{\alpha}}}
\left[\frac{1}{\overline{\omega}}\sum_{\lambda=\pm1}
(\lambda-\overline{\omega})L_{\lambda}+2\tan^{-1}\tilde{\alpha}\right],
\label{V-RD}
\\
&{\rm Im}D = \frac{3\pi}{32}\frac{\tilde{\alpha}^{-1}}{1+2\Delta_{\tilde{\alpha}}}
\frac{1}{\overline{\omega}}\sum_{\lambda}\lambda S_{\lambda}, 
\label{V-ID}\\
&{\rm Re}E_{1}=
\frac{2\Delta_{\tilde{\alpha}}\tilde{\alpha}^{-1}}{1+2\Delta_{\tilde{\alpha}}}
\frac{\epsilon_{\rm F}}{\hbar\omega}
+\frac{3}{8}\frac{\tilde{\alpha}^{-2}}{1+2\Delta_{\tilde{\alpha}}}
\left[
4\overline{\omega}\tan^{-1}\tilde{\alpha}
+2\sum_{\lambda}(\lambda-\overline{\omega})L_{\lambda}
+\frac{9}{8}\tilde{\alpha}^2\sum_{\lambda}\frac{\lambda-2\overline{\omega}}{\lambda-\overline{\omega}}
\lambda R_{\lambda}\right.
\nonumber\\
&\qquad~~\left.-\frac{3}{8}\tilde{\alpha}^2\sum_{\lambda}
\frac{(\lambda-3\overline{\omega})(1+\tilde{\alpha}^2R_{\lambda})}
{\overline{\omega}^2+2\lambda\tilde{\alpha}^2\overline{\omega}-\tilde{\alpha}^2}
\right]
,\label{V-RE1}\\
&{\rm Im}E_{1}=
\frac{3\pi}{32}\frac{\tilde{\alpha}^{-2}}{1+2\Delta_{\tilde{\alpha}}}
\left[\sum_{\lambda}\lambda S_{\lambda}
-\frac{15}{4}\overline{\omega}^2\sum_{\lambda}\frac{\lambda}{S_{\lambda}}
+\frac{3}{4}\overline{\omega}^2(2\tilde{\alpha}^4-\tilde{\alpha}^2-\overline{\omega}^2)
\sum_{\lambda}\frac{\lambda}{S_{\lambda}^3}
\right]
,\label{V-IE1}\\
&{\rm Re}E_{2}=
\frac{2\Delta_{\tilde{\alpha}}\tilde{\alpha}^{-1}}{1+2\Delta_{\tilde{\alpha}}}
\frac{\epsilon_{\rm F}}{\hbar\omega}-\frac{3}{8}\frac{\tilde{\alpha}^{-2}}{1+2\Delta_{\tilde{\alpha}}}
\left[
2\overline{\omega}\tan^{-1}\tilde{\alpha}
+\sum_{\lambda}(\lambda-\overline{\omega})L_{\lambda}
+\frac{3}{8}\tilde{\alpha}^2\sum_{\lambda}\frac{\lambda-2\overline{\omega}}{\lambda-\overline{\omega}}
\lambda R_{\lambda}\right.\nonumber\\
&\qquad~~\left.
-\frac{1}{8}\tilde{\alpha}^2\sum_{\lambda}
\frac{(\lambda-3\overline{\omega})(1+\tilde{\alpha}^2R_{\lambda})}
{\overline{\omega}^2+2\lambda\tilde{\alpha}^2\overline{\omega}-\tilde{\alpha}^2}
\right],\label{V-RE2}\\
&{\rm Im}E_{2}=
-\frac{3\pi}{32}\frac{\tilde{\alpha}^{-2}}{1+2\Delta_{\tilde{\alpha}}}
\left[\sum_{\lambda}\lambda S_{\lambda}
-\frac{5}{4}\overline{\omega}^2\sum_{\lambda}\frac{\lambda}{S_{\lambda}}
+\frac{1}{4}\overline{\omega}^2(2\tilde{\alpha}^4-\tilde{\alpha}^2-\overline{\omega}^2)
\sum_{\lambda}\frac{\lambda}{S_{\lambda}^3}
\right]
,\label{V-IE2}\\
&{\rm Re}E_{3}=
\frac{3}{8}\frac{\tilde{\alpha}^{-2}}{1+2\Delta_{\tilde{\alpha}}}
\left[
-2\overline{\omega}\tan^{-1}\tilde{\alpha}
-\sum_{\lambda}(\lambda-\overline{\omega})L_{\lambda}
-\tilde{\alpha}^2\sum_{\lambda}\frac{\lambda-2\overline{\omega}}{\lambda-\overline{\omega}}
\lambda R_{\lambda}
\right]
,\label{V-RE3}\\
&{\rm Im}E_{3}=
\frac{3\pi}{32}\frac{\tilde{\alpha}^{-2}}{1+2\Delta_{\tilde{\alpha}}}
2\overline{\omega}^2\sum_{\lambda}\frac{\lambda}{S_{\lambda}}
.\label{V-IE3}
\end{align}
Thus, from equations (\ref{V-IE1}) and (\ref{V-IE3}), 
we obtain the explicit form of ${\rm Im}E_{13}\equiv{\rm Im}(E_{1}+E_{3})$ 
as 
\begin{align}
{\rm Im}E_{13}
&=\frac{3\pi}{32}\frac{\tilde{\alpha}^{-2}}{1+2\Delta_{\tilde{\alpha}}}
\left[
\sum_{\lambda}\lambda S_{\lambda}
 -\frac{7}{4}\overline{\omega}^2\sum_{\lambda}\frac{\lambda}{S_{\lambda}}
 +\frac{3}{4}\overline{\omega}^2(2\tilde{\alpha}^4-\tilde{\alpha}^2-\overline{\omega}^2)
 \sum_{\lambda}\frac{\lambda}{S_{\lambda}^3}
\right]. 
\label{IE13}
\end{align}
We see that this function diverges towards the transition edges 
$\bar{\omega}_{\pm}$ because of the second and last terms.
The dominant term at the edges is the last term. 
Thus the directional dichroism is strongly enhanced (see main text). 
Note that similar equations as  (\ref{IS}) and (\ref{IN}) 
are understood in the equations (\ref{V-IE1}), (\ref{V-IE2}), (\ref{V-IE3}) and (\ref{IE13}).

\section{Toroidal and quadrupole moments}

In this section, we briefly make an analogy between our microscopic results 
of $\sigma_{ij}^{M}$ in (\ref{sigma-M}) 
and the conductivity induced by an effective toroidal and  quadrupole moments. 

Let us consider an effective Hamiltonian 
proposed to describe cross correlation effects: \cite{Spaldin08S} 
\begin{align}
H_{\rm d}= -g{\cal A}\cdot({\bm E}\times{\bm B})-hQ_{ij}E_{i}B_{j}
\end{align}
where $g$ and $h$ are coefficients, 
and ${\cal A}$ and $Q_{ij}(=Q_{ji})$ are a toroidal and a quadrupole moments, respectively. 
The induced current density is obtained from the derivative of the 
Hamiltonian with respect to the vector potential, ${\bm A}$, as 
\begin{align}
j_{{\rm d},i} &= -\frac{\delta H_{\rm d}}{\delta A_{i}}\equiv \sigma^{\rm d}_{ij}E_{j}, 
\end{align}
where the conductivity $\sigma^{\rm d}_{ij}$ is calculated as 
\begin{align}
\sigma^{\rm d}_{ij}= ig\omega\left[
2{\cal A}\cdot\qv\delta_{ij}-({\cal A}_{i}q_{j}+{\cal A}_{j}q_{i})
\right]
+
ih \omega\left(
Q_{il}\vare_{lkj}-\vare_{ikl}Q_{lj}
\right)q_{k}.
\end{align}

In the real space representation, the induced current is written as 
\begin{align}
{\bm j}_{\rm d}= \nabla \times \Mv_{\rm d}+\frac{\partial {\bm P}_{\rm d}}{\partial t},
\end{align}
where 
$M_{{\rm d},i} = -g({\cal A}\times {\bm E})_{i}- h Q_{ij}E_{j}$ and 
$P_{{\rm d},i} = g({\cal A}\times {\bm B})_{i} + h Q_{ij}B_{j}$ 
are an effective magnetization and electric polarization  
induced by the toroidal and quadrupole moments, respectively.

If we put respectively ${\cal A}$ and $Q_{ij}$ on 
${\cal A} = {\bm \alpha}\times\Mv$ and $Q_{ij} = \alpha_{i}M_{j} + \alpha_{j}M_{i}$, 
we have 
\begin{align}
\sigma_{ij}^{\rm d} &= 
2ig ({\bm \alpha}\times \Mv)\cdot\qv\delta_{ij}
-ig\left[
({\bm \alpha}\times \Mv)_{i}q_{j} + q_{j}({\bm \alpha}\times \Mv)_{i}
\right]
\nonumber\\
&+ih\left[M_{i}({\bm \alpha}\times \qv)_{j}+({\bm \alpha}\times\qv)_{i}M_{i}\right]
+ih\left[\alpha_{i}(\Mv\times \qv)_{i}+(\Mv\times\qv)_{i}\alpha_{j}\right]. 
\label{sigma-Q}
\end{align}
Thus, except for coefficients,  we can have derived the linear terms of 
$\sigma^{M}_{ij}$  with respect to $\qv$ and $\Mv$. 
Some differences between the two may come from the anisotropic property due to the Rashba field ${\bm \alpha}$. 

\end{document}